\documentclass[amsmath, amssymb, aps, showpacs, prd, twocolumn, nofootinbib, superscriptaddress]{revtex4-1}
\usepackage[T1]{fontenc}
\usepackage{ae,aecompl}
\usepackage{enumerate, enumitem}
\usepackage[colorlinks=true, linkcolor=blue, citecolor=red]{hyperref}
\usepackage{graphicx}
\usepackage{color}
\usepackage{multirow}
\usepackage[normalem]{ulem}
\usepackage{mathtools}
\usepackage[justification=centerlast, singlelinecheck=false, caption=false]{subfig}

\renewcommand{\vec}{\boldsymbol}
\newcommand{\im}{\mathrm{i}}
\newcommand{\rd}{\mathrm{d}}
\newcommand{\re}{\mathrm{e}}

\newcommand{\mixInd}[3]{{#1}^{#2}_{\hphantom{#2}#3}}
\newcommand{\T}{\prescript{1}{}{T}}

\setlist[itemize]{leftmargin=*}

\begin{document}

\title{Brane induced gravity: Ghosts and naturalness}

\author{Ludwig Eglseer}
\email[]{l.eglseer@physik.uni-muenchen.de}
\affiliation{Arnold Sommerfeld Center for Theoretical Physics, Ludwig-Maximilians-Universit\"at, Theresienstra{\ss}e 37, 80333 Munich, Germany}

\author{Florian Niedermann}
\email[]{florian.niedermann@physik.uni-muenchen.de}
\affiliation{Arnold Sommerfeld Center for Theoretical Physics, Ludwig-Maximilians-Universit\"at, Theresienstra{\ss}e 37, 80333 Munich, Germany}

\author{Robert Schneider}
\email[]{robert.bob.schneider@physik.uni-muenchen.de}
\affiliation{Arnold Sommerfeld Center for Theoretical Physics, Ludwig-Maximilians-Universit\"at, Theresienstra{\ss}e 37, 80333 Munich, Germany}
\affiliation{Excellence Cluster Universe, Boltzmannstra{\ss}e 2, 85748 Garching, Germany\\
~\\}

\date{\today}

\begin{abstract}
Linear stability of brane induced gravity in two codimensions on a static pure tension background is investigated. The brane is regularized as a ring of finite circumference in extra space. By explicitly calculating the vacuum persistence amplitude of the corresponding quantum theory, we show that the parameter space is divided into two regions---one corresponding to a stable Minkowski vacuum on the brane and one being plagued by ghost instabilities. This analytical result affirms a recent nonlinear, but mainly numerical analysis.

The main result is that the ghost is absent for a sufficiently large brane tension, in perfect agreement with a value expected from a natural effective field theory point of view. Unfortunately, the linearly stable parameter regime is either ruled out phenomenologically or becomes unstable for nontrivial cosmologies. We argue that super-critical brane backgrounds constitute the remaining window of opportunity.

In the special case of a tensionless brane, we find that the ghost exists for all phenomenologically relevant values of the induced gravity scale. Regarding this case, there are contradicting results in the literature, and we are able to fully resolve this controversy by explicitly uncovering the errors made in the ``no-ghost'' analysis. Finally, a Hamiltonian analysis generalizes the ghost result to more than two codimensions.
\end{abstract}


\maketitle

\section{Introduction}

General Relativity (GR) viewed as an effective field theory (EFT) gives rise to a famous problem: the observed value of the cosmological constant is numerically not stable under quantum corrections---an observation which is usually referred to as the cosmological constant problem~\cite{Weinberg:1988cp}. An auspicious way of addressing this problem relies on modifying GR by weakening gravity on cosmological scales and thereby making it insensitive to a cosmological constant (for recent reviews of modified gravity see~\cite{Joyce:2014kja, deRham:2014zqa}, and for the so called degravitation mechanism see~\cite{Dvali:2002pe, Dvali:2002fz, Dvali:2007kt, deRham:2007rw, Burgess:2011va}). 

A prominent arena for realizing such an infrared modification is provided by the braneworld scenario (see~\cite{Maartens:2010ar} for a review), according to which our universe is a surface---the brane---in some higher dimensional spacetime---the bulk. The crucial idea then is to absorb the gravitational impact of a cosmological constant into extrinsic curvature, invisible to a brane observer~\cite{Rubakov:1983bz,Charmousis:2001hg, ArkaniHamed:2002fu}.  In principle, such a surface can be realized by a topological defect like a domain wall in five, or a vortex in six dimensions. In both cases there are known ways of localizing particles inside the defect, see \cite{Rubakov:1983bb} and \cite{Dvali:2001ae}, respectively. These constructions generically introduce a length scale $r_0$ that characterizes the transverse width of the defect. However, as long as we are interested in low energy questions, which take place on length scales $\ell$ much larger than the defect size, it is not necessary to resolve its microscopic details. In particular, for cosmological applications where $\ell$ is of the order of the Hubble length, $\ell \gg r_0$  by at least 30 orders of magnitude. Thus, in this context the defect can in good approximation be treated as infinitely thin.\footnote{In higher codimensions, delta sources generically lead to divergences, and therefore have to be regularized, see Sec.~\ref{sec:regularization}. However, as we will see in Sec.~\ref{sec:hamiltonian}, some questions can already be addressed at the zero-width level.} Then, in accordance with the EFT paradigm, it suffices to write down all operators that are compatible with the symmetries of the theory. In the case at hand this is the higher dimensional diffeomorphism invariance in the bulk and its four dimensional counterpart on the brane. As long as we do not impose any further symmetries (like supersymmetry~\cite{Burgess:2011va}), we find
\begin{align}  
	\label{eq:ActionBIG}
	\mathcal{S}=
	\mathcal{S}_{\rm EH}+
	\mathcal{S}_{\rm BIG}+
	\mathcal{S}_{\rm m}\;.
\end{align}
The first term,
\begin{align}
	\mathcal{S}_{\rm EH}=
	M_{D}^{D-2}\int {\rm d}^D X\;\sqrt{-g}\left( \mathcal{R}^{(D)}-2\lambda^{(D)} \right)\,,
\end{align}
describes Einstein-Hilbert gravity in $ D = 4 + n $ infinite spacetime dimensions, where $ \mathcal{R}^{(D)} $ is the $ D $-dimensional Ricci scalar, the bulk Planck scale is denoted by $M_D$ and the bulk cosmological constant by $\lambda^{(D)}$.
The second term is the induced gravity term on a codimension-$n$ brane:
\begin{align}
	\label{eq:S_BIG_delta_0}
	\mathcal{S}_{\rm BIG}= M_4^2\int {\rm d}^4 x\,\sqrt{- g^{(4)}}\; \left(\mathcal{R}^{(4)}-2 \lambda^{(4)}\right) \,.
\end{align}
The induced terms are controlled by the induced gravity scale\footnote{For phenomenological reasons, $ M_4 $ should be identified with the Planck mass.} $M_4$ and the brane tension $\lambda:=M_4^2  \lambda^{(4)}$. The last term in \eqref{eq:ActionBIG} is the action for matter fields localized on the brane and coupled to the induced metric $g^{(4)}_{\mu\nu}$; ultimately, it contains all Standard Model fields. In its pure form and for infinite volume extra dimensions, this theory is known as Brane Induced Gravity (BIG) in general, or the Dvali-Gabadadze-Porrati (DGP) model in the special case of $n=1$  \cite{Dvali:2000hr,Dvali:2000xg}. It is important to note that even if we tuned the brane induced parameters $M_4$ and  $\lambda$ to zero at the classical level, they would be generated nonetheless at the quantum level by particle loops on the brane. In other words, irrespective of the underlying microscopic theory that gave rise to the surface theory, the BIG terms have to be included in order to meet the requirements of a natural EFT.

That being said, we arrive at the following puzzle: There are several claims in the literature~\cite{Dubovsky:2002jm, Hassan:2010ys} stating that for $\lambda^{(D)}=\lambda=0$ and $ n \geq 2 $ the theory \eqref{eq:ActionBIG} has a tachyonic ghost in a weakly coupling regime on a Minkowski background, thus raising the question: \textit{Is the EFT description of gravity induced on higher codimensional surfaces necessarily plagued by instabilities?} This question is at the core of the present work. An answer was supposedly given in a recent work~\cite{Berkhahn:2012wg}, which claimed that the theory is healthy in that parameter regime ($\lambda=0$), in contradiction to the former results. 

By performing two independent analyses, we are able to disprove this claim of~\cite{Berkhahn:2012wg} and to confirm the tachyon and ghost result from the literature~\cite{Dubovsky:2002jm, Hassan:2010ys} for arbitrary codimensions $ \geq 2 $. Moreover, we are able to pin down the errors that were made in \cite{Berkhahn:2012wg} by explicitly correcting the old calculations here in Appendix~\ref{app:errata} and fully reconciling them with our new analysis.

However, this is not the end of the story. In \cite{Kaloper:2007ap} and \cite{Niedermann:2014bqa} the parameter range was extended to non-zero but sub-critical\footnote{A pure tension codimension-two brane is known to produce a conical geometry characterized by a deficit angle in extra space. For the critical value of the tension, $ \lambda_\mathrm{c} \equiv 2\pi M_6^4 $, it becomes $2\pi$ leading to a cylindrical geometry, cf.\ \cite{Linet:1990fk} or \cite{Niedermann:2014bqa} for a more recent discussion. Even larger (super-critical) values lead to a different topology and inflating branes~\cite{Niedermann:2014yka}; they are excluded in the present work.} values of the brane tension $\lambda$ (but keeping for simplicity $\lambda^{(D)}=0)$. In  \cite{Kaloper:2007ap} a linear perturbation analysis was performed and a particular scalar mode of the dynamical spectrum of the theory was shown to lead to a strong coupling for phenomenologically interesting parameter values, which only vanished in a near-critical tension regime. In \cite{Niedermann:2014bqa} a completely nonlinear but numerical analysis for homogeneous and isotropic matter on the brane was performed, showing the existence of both a stable and unstable parameter regime. Unfortunately, the stable regime had to be ruled out on phenomenological grounds. It should be noted that the two works are based on slightly different physical assumptions: While the analysis in~\cite{Kaloper:2007ap} allowed for small fluctuations of the circumference of the brane described by the radion field, the one in~\cite{Niedermann:2014bqa} assumed a strictly constant circumference, or an infinitely heavy radion equivalently, based on some underlying stabilization mechanism. To summarize, the results of \cite{Kaloper:2007ap} pointed towards the near-critical regime as a remaining window of opportunity, whereas the cosmological analysis of \cite{Niedermann:2014bqa} excluded all sub-critical parameter space but lacked a completely analytic statement. 

The analysis of this work will establish a comprehensive and \textit{analytic} picture which nicely confirms the results of \cite{Dubovsky:2002jm, Hassan:2010ys,Niedermann:2014bqa} and is fully compatible with \cite{Kaloper:2007ap}. To be more specific, by focusing on the codimension-two model, we solve the full system of bulk vacuum and brane matching equations at the linear level\footnote{In general, the linear analysis is sufficient for addressing weak field questions like e.g.~solar system predictions. Note however that for certain parameter choices there might emerge a Vainshtein-like regime around massive sources \cite{Vainshtein:1972sx}. Therefore, in this work we limit ourselves to sources that can be described within a weakly coupled theory. Moreover, for cosmological questions one needs to resort to a nonlinear analysis as in~\cite{Niedermann:2014bqa}.} around a conical background geometry introduced in Sec.~\ref{sec:background}. To that end, we regularize the setup by wrapping the brane around a circle in extra space and thus promoting the brane to a codimension-one object. The gravitational effect of the brane tension is then completely absorbed by the conical defect, while the geometry on the brane is Riemann flat. Thus, from the perspective of a brane observer there is a landscape of different vacua corresponding to different values of $\lambda$~\cite{Kaloper:2007ap}. (For that reason these solutions are interesting with respect to the cosmological constant problem, as the brane tension plays the role of a 4D cosmological constant.) Like in the approach of \cite{Niedermann:2014bqa}, we assume the existence of a stabilization mechanism in the microscopic theory that forces the compact brane dimension to be of constant size. On the level of fluctuations, this corresponds to setting the radion field to zero, and can be achieved by appropriately dialing the angular pressure fluctuations. (In Appendix~\ref{ap:non_stabil} we drop that assumption. Accordingly, the radion field represents an additional dynamical degree of freedom. Moreover, in Appendix~\ref{ap:radion_stabil} we provide an explicit example of a dynamical stabilization mechanism.)

In Sec.~\ref{sec:lin_stab} we introduce the vacuum persistence amplitude $  \langle 0 | 0 \rangle_T  $ as a diagnostic tool, which enables us to probe for the presence of a ghost in the fluctuation theory. This is based on the fact that a ghost mode, treated with the standard Feynman prescription, would lead to a transition probability $ \left| \langle 0 | 0 \rangle_T \right|^2 $ larger than one.

A Hamiltonian analysis in Sec.~\ref{sec:hamiltonian} offers a different but compatible perspective and generalizes the results to arbitrary codimensions for vanishing $\lambda$. It also enables us to do a solid counting of degrees of freedom in arbitrary dimensions. Furthermore, it corrects the erroneous expression for the Hamiltonian derived in \cite{Berkhahn:2012wg}. The final discussion in Sec.~\ref{sec:discussion} confirms the consistency of \eqref{eq:ActionBIG} as an EFT and thereby answers the initiatory question with ``no'': \emph{the instability in codimension two BIG disappears if the brane tension is not tuned unnaturally small}.

\subsection*{Summary}

To summarize, our investigations lead to the following main findings:
\begin{itemize}
\item We derive a function $f(M_6,M_4, \lambda, r_0) $ of the model parameters, the sign of which determines the healthy (stable and ghost-free) and pathological (unstable and ghostly) regions in the (sub-critical) parameter space. This function is the static version of the one found in the cosmology analysis of~\cite{Niedermann:2014bqa}. 

\item For a natural choice of brane induced parameters the theory is in the healthy regime. To be specific, there are \emph{stable} on-brane Minkowski vacua, for which the tension is completely absorbed into extrinsic curvature. The instability only emerges if the induced gravity scale $M_4$ is large while the tension is kept small.  

\item The stable vacua are ruled out phenomenologically except for the near-critical regime. However, the cosmological analysis in~\cite{Niedermann:2014bqa} showed that for FLRW symmetries on the brane, i.e.\ a background geometry with non-vanishing 4D Hubble parameter $H$, the stable regime becomes significantly smaller, thereby closing the remaining window of opportunity (at least for sub-critical brane tensions).

%However, this regime suffers from an instability that is induced by higher order corrections as can be inferred from the nonlinear analysis in \cite{Niedermann:2014bqa}.

\item The ghost found in \cite{Dubovsky:2002jm, Hassan:2010ys}, the strong coupling observed around static sources in \cite{Kaloper:2007ap} and the instability encountered in \cite{Niedermann:2014bqa} is produced by the same scalar degree of freedom $s$ and thus the manifestation of the same pathology: for model parameters with $f>0$ the scalar mode $s$ is a tachyonic ghost with mass $m_*$.  A priori, the status of the ghost is not clear as it could be an artifact of the unstable conical geometry. Indeed, the background is destabilized due to the tachyonic character of the scalar $s$, which leads to exponentially growing low momentum modes characterized by  $|\vec{p}|<m_*$. However, the nonlinear analysis of~\cite{Niedermann:2014bqa} did not show any indication for the formation of a new stable background, instead the solutions exhibited a run-away behavior. 

\item Higher codimensional versions share the same pathology, at least on the parameter subspace characterized by $\lambda=0$, in accordance with~\cite{Dubovsky:2002jm, Hassan:2010ys}.

\item For more than one codimension, the theory has six gravitational degrees of freedom that couple to on-brane sources and are thus phenomenologically relevant\footnote{If the regularized brane width is not stabilized, there is an additional degree of freedom corresponding to fluctuations of its angular size.}. The total number of degrees of freedom is found to be $D(D-3)/2$ which is the same as in Einstein gravity.

\item Without external stabilization, the radion is tachyonic in a broad (and relevant) parameter regime; it thus destabilizes the background geometry. In other words, assuming some sort of underlying modulus stabilization is required for having a consistent regularization.
\end{itemize}

\subsection*{Conventions}

As our sign convention we adopt ``$ +++ $'' as defined in~\cite{Misner}. We work in units in which $ c = \hbar = 1 $. While we work in six dimensions in Secs.~\ref{sec:background} and~\ref{sec:lin_stab}, we generalize the spacetime dimensions to $D=4+n$, where $n$ denotes the number of codimensions, in Sec.~\ref{sec:hamiltonian} and Appendix~\ref{app:errata}. The notational conventions for the index ranges are defined according to:
\begin{table}[ht]
\centering
\begin{tabular}{ |l |l c| }
\hline
$ A, B, \ldots $ &  \multicolumn{1}{l |}{$0$, $1$, $2$, $3$, $r$, $\phi$} & \multirow{ 2}{*}{(in Secs.~\ref{sec:background} \& \ref{sec:lin_stab})}\\
\cline{1-2}
$ a, b $ & \multicolumn{1}{l |}{$r$, $\phi$}& \\
\hline
$ A, B, \ldots $ & \multicolumn{1}{l |}{$0$, $1$, $2$, $3$, $5$, $6$ \ldots,  $D$ }& (in Sec.~\ref{sec:hamiltonian} \& \\
\cline{1-2}
$ a, b, \ldots $ & \multicolumn{1}{l |}{$5$, $6$, \ldots, $D$ }&  Appendix~\ref{app:errata}) \\
\hline
$ \alpha, \beta, \gamma, \delta $ &$0$, $1$, $2$, $3$,  $\phi$& \\
\hline
$\mu, \nu, \rho, \sigma $ & $0$, $1$, $2$, $3$ & \\
\hline
$ i, j, \ldots $ & $1$, $2$, $3$ & \\
\hline
\end{tabular}
\end{table}

\noindent
The index values $ r $ and $ \phi $ are used to emphasize the use of polar coordinates. For symmetrization we use the convention $\partial_{(i} V_{j)}=\left(\partial_{i} V_{j}+\partial_{j} V_{i}\right)/2$. Four-momentum is denoted by $ p = (\omega, \vec{p}) $.

\section{Cosmic string background}
\label{sec:background}

In this section, we introduce the ring regularization of the codimension-two brane and present the corresponding equations of motion, as well as the static deficit angle background that is generated by a pure-tension brane.

\subsection{Regularization} \label{sec:regularization}
%By analogy to a charged string in electrostatics, we expect the gravitational field to diverge logarithmically at the position of a codimension-two defect. 
Just like for a charged string in electrostatics, the gravitational field in general diverges logarithmically at the position of a codimension-two defect.
This problem can be taken care of by introducing a transverse width of the brane. From a more fundamental perspective we could have done this right from the beginning because there has to be a physical mechanism which creates the brane in the first place. Topological defects---as for example a Nielsen-Olesen vortex~\cite{Nielsen:1973cs} in two codimensions---would be obvious candidates. In general, these solutions predict a finite transverse spread of the brane.

Instead, in favor of technical simplicity, we choose to describe the brane as a ring $ \mathcal{S}_1 $ of circumference $2\pi r_0$ in extra space. Hence, it becomes a codimension-one object, separating an interior from an exterior vacuum region. This regularization has already proved to be convenient in earlier works \cite{Peloso:2006cq, Kaloper:2007ap, Niedermann:2014bqa}. Technically, it amounts to the replacement\footnote{The origin as a codimension-two model requires the existence of a regular axis in the interior region, which will be demanded throughout this work. Otherwise, the same action could also describe a topologically distinct model without an axis~\cite{Niedermann:2014vaa}. }
\begin{equation}
\label{reg_sub}
\mathcal{S}_{\rm BIG} ~\longrightarrow~ M_5^3\int_{\mathcal{M}_4 \times \mathcal{S}_1}\!\!\!\!\! {\rm d}^5 x\,\sqrt{- g^{(5)}}\left( \mathcal{R}^{(5)}-2\lambda^{(5)}\right)	\;,
\end{equation}  
where $ M_5^3 = M_4^2 / (2 \pi r_0) $, $\lambda^{(5)}=\lambda/(2\pi r_0 M_5^3)$ and $g^{(5)}_{\alpha\beta}$ is the five dimensional induced metric. Of course, at this stage the process of introducing a particular regularization is arbitrary. Instead, we could have smeared the field over a disc of radius $r_0$ or similarly used a blurring function \cite{Hassan:2010ys}. We could have even resolved the brane in more fundamental degrees of freedom by using the aforementioned Nielsen-Olesen construction. However, the crucial point is that we limit the range of applicability of the theory to scales $\ell \gg r_0$. In other words, as long as we are interested in low energy questions, as they arise for example in the context of late time cosmology, we do not have to resolve the microscopic brane physics; in particular, we expect low energy questions not to depend on the chosen regularization scheme. 

In order to have a consistent regularization, we need to stabilize the circumference of the ring. Effectively, this can be realized by imposing a certain (non-constant) amount of external pressure in angular direction which prevents the circumference of the brane from fluctuating. In a fundamental picture, this corresponds to a model where the angular mode, describing fluctuations of the circumference, is very heavy compared to the typical energy scale $\ell^{-1}$. In Appendix~\ref{ap:radion_stabil} we discuss an explicit example of such a dynamical stabilization mechanism and show that our effective method is appropriate.

In that context, let us note that there is a lower bound on the mass of the angular mode (or radion)---as well as the inverse regularization width $ 1/r_0 $---on the basis of post Cavendish experiments which is $\sim10^{-3} {\rm eV}$, corresponding to a Compton wavelength of $\sim 100 \mu{\rm m}$ (for a recent work, see \cite{Kapner:2006si}, for a review see~\cite{Joyce:2014kja} and references therein). This means that for cosmological applications, corresponding to much smaller energies, it is fully justified to ignore those angular size fluctuations by setting the radion to zero or equivalently its mass to infinity.

However, in order to show that our crucial results are independent of that assumption, we also study the case where the size modulus is light and thus can be excited as an additional degree of freedom, cf.\  Appendix~\ref{ap:non_stabil}. As a result, we find that the background geometry gets destabilized as the circumference starts to grow (or decrease) exponentially in a broad parameter regime. This shows that the stabilization requirement is indeed necessary to avoid additional instabilities. The crucial point, however, is that the ghost survives in this case, and thus is no artifact of the ring stabilization.

By performing a Hamiltonian analysis in Sec.~\ref{sec:hamiltonian} (and a Lorentz covariant analysis in Appendix \ref{app:cov_analysis}), we can explicitly confirm that the ghost result does not depend on the details of the regularization. In that case we do not need to solve the equations explicitly and can therefore stick to the codimension-two description in terms of delta functions. This is an important and successful consistency check of our main analysis, which relies on a specific regularization.

\subsection{Equations of motion}

Since the bulk is source-free, the six-dimensional Einstein tensor has to satisfy the vacuum field equations away from the brane ($ r \neq r_0 $),
\begin{equation}\label{eq:vac_einstein}
	G^{(6)}_{AB} = 0 \,.
\end{equation}
These have to be supplemented by Israel's junction conditions~\cite{Israel:1966, Israel:1967} across the brane, including the BIG terms,
\begin{equation}\label{eq:israel}
	M_6^4 \big([K^\gamma_{\hphantom{\gamma}\gamma}] \delta^{\alpha}_{\hphantom{\alpha}\beta} - [K^{\alpha}_{\hphantom{\alpha}\beta}]\big) + M_5^3 G^{(5)\alpha}_{\hphantom{(5)\alpha}\beta} = T^{(5)\alpha}_{\hphantom{(5)\alpha}\beta} \,,
\end{equation}
where $ K^{\alpha}_{\hphantom{\alpha}\beta} $ is the extrinsic curvature, and square brackets denote the jump across the brane, i.e.\ $ [f] := f_{\mathrm{out}} - f_{\mathrm{in}} $.

%Note that coordinates in the interior and exterior can be chosen completely independently; only the metric has to be continuous across the brane. Therefore, the vacuum equations~\eqref{eq:vac_einstein} should actually be read as two independent sets of equations, one inside and one outside the brane.

\subsection{Deficit angle solution}
\label{sec:def_angle_sol}
For a pure tension brane the source reads
\begin{equation}
	\label{eq:EMT_4D_tension}
	T^{(5)\alpha}_{\hphantom{(5)\alpha}\beta} = - \frac{\lambda}{2\pi r_0} \delta^{\alpha}_{\hphantom{\alpha}\mu} \delta^{\mu}_{\hphantom{\mu}\beta} \,, 
\end{equation}
where the factor $ 2\pi r_0 $ is chosen such that $ \lambda $ corresponds to a 4D brane tension. For the static solution to exist, the pressure in angular direction was dialed to zero. In Appendix~\ref{ap:radion_stabil} we provide an example how this can be achieved microscopically. The exact solution is well known (see e.g.\ \cite{Niedermann:2014bqa} and references therein), and can be written in the form:
\begin{subequations} \label{eq:def_angle_sol}
\begin{align}
	\rd s^2 & = \eta_{\mu\nu} \rd x^{\mu} \rd x^{\nu} + \rd r^2 + g(r)^2 \rd \phi^2 \\
	g(r) &:= 
	\begin{cases}
		r & (r < r_0) \\
		r_0 + \left( 1 - \frac{\delta}{2\pi} \right) (r - r_0) & (r > r_0)
	\end{cases}
\end{align}
\end{subequations}
Here we used a continuous coordinate patch with polar coordinates $ (r, \phi) $ in the two extra dimensions, having the standard ranges $ r \in [0, \infty) $, $ \phi \in [0, 2\pi) $. The brane is located at $ r = r_0 $.
The geometrical meaning of the solution~\eqref{eq:def_angle_sol} is that (i) the 4D on-brane geometry is completely flat and (ii) the 6D geometry is locally flat as well, but has a deficit angle $ \delta $ in the exterior, which is related to the brane tension by
\begin{equation}
	\delta = \frac{\lambda}{M_6^4} \,.
\end{equation}
In this paper we will only consider sub-critical tensions, i.e.\ $ \lambda < 2\pi M_6^4 \Leftrightarrow \delta < 2\pi $. Otherwise, the static deficit angle solution would not be stable, the on-brane geometry would instead inflate and the bulk topology would be different~\cite{Niedermann:2014yka}.

\section{Linear stability analysis} \label{sec:lin_stab}

We will now consider small metric perturbations around the deficit angle background~\eqref{eq:def_angle_sol},
\begin{equation}
	g_{AB} = \gamma_{AB} + h_{AB} \,,
\end{equation}
where
\begin{equation} \label{eq:met_bkg}
	\gamma_{AB} = \mathrm{diag} \left( -1, 1, 1, 1, 1, g(r)^2 \right) \,,
\end{equation}
i.e.\ we still work in polar coordinates $ X^A = \left ( x^{\mu}, r, \phi \right ) $. Note that all indices on first order quantities will be lowered and raised with the background metric $ \gamma_{AB} $ and its inverse $ \gamma^{AB} $.

The question we want to answer is whether $ h_{AB} $ contains---at the linear level---instable modes (tachyons or ghosts, or both), which can be sourced by an additional (small) on-brane source $ \mixInd{U}{\alpha}{\beta} $, i.e.
\begin{subequations}
\begin{align}
	\mixInd{T}{(5)\alpha}{\beta} & = \prescript{0}{}{T}^{(5)\alpha}_{\hphantom{(5)\alpha}\beta} + \T^{(5)\alpha}_{\hphantom{(5)\alpha}\beta}\\
	& =: -\frac{\lambda}{2\pi r_0} \delta^{\alpha}_{\hphantom{\alpha}\mu} \delta^{\mu}_{\hphantom{\mu}\beta} - \frac{1}{2} \mixInd{U}{\alpha}{\beta} \,.
\end{align}
\end{subequations}
Since the source is distributed in a $ \phi $-symmetric way, it is sufficient to consider metric perturbations that also respect this symmetry, because only those are sourced. This means that the $ h_{AB} $ are $ \phi $-independent functions, and that $ h_{\phi\mu} = h_{\phi r} = 0 $. Furthermore, we can keep the brane at the fixed coordinate position $ r_0 = \mathrm{const} $ without loss of generality,  because a proper motion in radial direction (which is not ruled out by stabilizing the $ \phi $-direction), as well as a dependence of the physical brane radius on the spatial brane coordinates, can still be accomplished by allowing for nonzero $ h_{\mu r} $ components.
The metric perturbations therefore take the form
\begin{equation} \label{eq:split_3D_0}
	h_{AB} =
	\begin{pmatrix}
		h_{\mu\nu} & h_{\mu r} & 0 \\
		h_{r\nu} & h_{rr} & 0 \\
		0 & 0 & h_{\phi\phi}
	\end{pmatrix} =:
	\begin{pmatrix}
		-N & h_{0j} & l' & 0 \\
		h_{i0} & h_{ij} & h_{ir} & 0 \\
		l' & h_{rj} & h_{rr} & 0 \\
		0 & 0 & 0 & h_{\phi\phi}
	\end{pmatrix} \,.
\end{equation}
Here and henceforth a prime is shorthand notation for $ \partial_r $.
It is convenient to decompose the 3D spatial components of $ h_{AB} $ as
\begin{subequations} \label{eq:split_3D}
\begin{align}
	h_{0i} & = N_i + \partial_i L \,, \label{eq:split_0i}\\
	h_{ij} & = D_{ij} + \partial_{(i} V_{j)} + \partial_i \partial_j B + \delta_{ij} S \,,\label{eq:split_ij}\\
	h_{ir} & = G_i' + \partial_i H' \,, \label{eq:split_ri}
\end{align}
\end{subequations}
where---from a 3D point of view---$ D_{ij} $ is a transverse traceless tensor, and $ N_i, V_i, G_i $ are transverse vectors, i.e.
\begin{subequations}
\begin{align}
	D^i_{\hphantom{i}i} = \partial_i D^i_{\hphantom{i}j} &= 0 \,,\\
	\partial_i N^i = \partial_i V^i = \partial_i G^i &= 0 \,,
\end{align}
\end{subequations}
while $ N, l, L, B, S, H $ are scalars.
Even though this decomposition is not manifestly Lorentz-covariant like the approach in~\cite{Berkhahn:2012wg}, it has the great advantage that it is invertible on the space of bounded functions, and so it does not introduce any ``split ambiguity'' (cf.\ Appendix~\ref{app:cov_analysis}). The reason is of course that the Laplace operator $ \Delta_3 $ has no non-trivial bounded solutions, unlike the d'Alembert operator $ \Box_4 $. This makes the identification of dynamical degrees of freedom much more straightforward.

The analogous decomposition of the $ (ab) $-components is\footnote{Note that in codimension two there is no tensor part, and the vector part is absent due to $ \phi $-symmetry.}
\begin{equation} \label{eq:def_b_s}
	h_{ab} = \nabla_{a} \nabla_{b} b + \gamma_{ab} s \,,
\end{equation}
with $ \nabla_a $ denoting the covariant derivative with respect to the background metric $ \gamma_{ab} $. Explicitly, this gives
\begin{align} \label{eq:h_b_s}
	h^r_{\hphantom{r}r} = b'' + s\,, && h^\phi_{\hphantom{\phi}\phi} = \frac{g'}{g} b' + s \,.
\end{align}
Again, the relation~\eqref{eq:def_b_s} is invertible, because the Laplace operator $ \Delta_2 $ has an empty kernel.

\subsection{Gauge-invariant variables}

The linearized bulk theory is invariant under the gauge transformations
\begin{equation} \label{eq:gauge}
	\delta h_{AB} = \nabla_{(A} \xi_{B)} \,.
\end{equation}
In order not to spoil the $ \phi $-symmetry, the $ \xi_A $ are subject to
\begin{align} \label{eq:gauge_symm}
	\xi_\phi = 0 \,, && \partial_\phi \xi_\mu = \partial_\phi \xi_r = 0 \,.
\end{align}
Instead of choosing a particular gauge, we will work with a complete set of gauge-invariant variables, which can chosen to be\footnote{One might worry that, since the definitions of $ P, Q $ and $ C_i $ involve time derivatives of some metric functions, one could be turning actual dynamical quantities into constrained ones ``by hand''. However, this is not the case, as can explicitly be seen by choosing the gauge $ B = b = L = G_i = 0 $ (in the bulk), which is always possible.}
\begin{subequations} \label{eq:g_inv_var_3D}
\begin{align}
	& D_{ij}\,, & & s\,, \\
	J &:= 3 S + s\,, & O & := B + b - 2H\,, \\
	P &:= \dot B - \dot b - 2 (L - l) \,, & Q & := \ddot B - N - 2\dot L\,, \label{eq:def_P_Q}\\
	C_i &:= N_i - \dot G_i \,, & W_i &:= 2 G_i - V_i\,,
\end{align}
\end{subequations}
where we introduced the dot as shorthand for $ \partial_t $.

Since we chose coordinates in which the brane is located at a fixed coordinate position ($ r = r_0 $), there is a further on-brane restriction on the gauge transformations,
\begin{equation}
	\xi_r|_0 = 0 \,,
\end{equation}
where the subscript ``0'' denotes evaluation at the brane. This implies that on the brane, there exists an additional gauge invariant function, namely
\begin{equation}
	\varphi := h^\phi_{\hphantom{\phi}\phi} |_0 \,,
\end{equation}
which will also appear explicitly in the junction conditions below.
Physically, it corresponds to the radion field, measuring fluctuations in the size-modulus of the regularized brane.

\subsection{Bulk equations of motion}

The bulk vacuum field equations~\eqref{eq:vac_einstein} at linear order in $ h_{AB} $ read
\begin{multline}\label{eq:vac_einstein_lin}
	\Box_6 h_{AB} + \nabla_A \nabla_B h^C_{\hphantom{C}C} - 2 \nabla_C \nabla_{(A} h^C_{\hphantom{C}B)} \\
	+ \gamma_{AB} \left( \nabla^C\nabla^D h_{CD} - \Box_6 h^C_{\hphantom{C}C} \right) = 0 \,,
\end{multline}
where $ \Box_6 := \nabla^A \nabla_A $.
These can now be projected onto the tensor, vector and scalar components, according to the decomposition of the metric perturbations~\eqref{eq:split_3D}. Let us emphasize that this projection only requires to divide by Laplace operators, which does not introduce any homogeneous functions in the resulting equations. This is in contrast to the 4D covariant split adopted in~\cite{Berkhahn:2012wg}, where one has to divide by d'Alembert operators, making the analysis much more subtle and complicated, see Appendix~\ref{app:cov_analysis}.

In the following, we omit all equations which are redundant due to the Bianchi identities. However, since we are particularly interested in distinguishing dynamical from constrained degrees of freedom, we only omit those components of~\eqref{eq:vac_einstein_lin} which appear in the Bianchi identities without time-derivatives. In other words, we always keep the stronger equations. Explicitly, we drop the $ (ij)^{(V)} $, $ (0i)^{(L)} $, $ (ir)^{(H)} $ and $ (rr) $ equations.

The resulting complete set of bulk equations of motion, expressed in terms of the gauge invariant variables~\eqref{eq:g_inv_var_3D} is:

\begin{itemize} %[leftmargin=*]
\item Tensor:
\begin{equation}
	\Box_6 D_{ij} = 0
\end{equation}
This is simply the $ (ij) $-component of~\eqref{eq:vac_einstein_lin}, projected onto the tensor part. It shows that $ D_{ij} $ is dynamical, carrying two independent degrees of freedom (DOF).

\item Vector:
\begin{subequations}
\begin{align}
	\Delta_3 W_i + 2 \dot C_i & = 0 \label{eq:bulk_vector_constr}\\
	\Box_6 C_i & = 0
\end{align}
\end{subequations}
The first one is the vector-projected $ (ir) $-bulk equation, showing that $ W_i $ is constrained.
The second one is the vector-projection of the $ (0i) $-bulk equation, with $ W_i $ eliminated by means of the constraint. Thus, $ C_i $ is dynamical, carrying two DOF.

\item Scalar:
\begin{subequations}\label{eq:constr_system}
\begin{align}
	\left (2 \Delta_3 + 3 \Delta_2 \right ) J + 4 \Delta_3 s + 3 \Delta_2\Delta_3 O & = 0 \label{eq_bulk_scalar_c} \\
	2 \dot J + \Delta_3 \left ( \dot O + P \right ) & = 0 \label{eq_bulk_scalar_P}\\
%	\left ( 3 \Box_4 - \Delta_3 \right ) J + \Delta_3 s - 3 \Delta_3 Q & = 0 \\
	J - s - Q + \Delta_3 O + \dot P & = 0 \label{eq:bulk_scalar_3}\\
	\Box_6 J = 0 \,, \quad \Box_6 s & = 0 \label{eq:bulk_scalar_dyn}
\end{align}
\end{subequations}
The first equation is the $ (00) $-component of~\eqref{eq:vac_einstein_lin}, the second one is its $ (0r) $-component (already integrated once in $ r $, requiring fall-off conditions in the bulk) and the third one is the difference of the $ (\phi\phi) $- and the $ (rr) $-equations (also integrated in $ r $). These three are constraint equations that can---for instance---be solved for $ O, P $ and $ Q $.
Plugging these solutions into the two scalar-projected $ (ij) $-components of~\eqref{eq:vac_einstein_lin}, and taking suitable linear combinations, yields the two dynamical equations~\eqref{eq:bulk_scalar_dyn} for $ J $ and $ s $.

\end{itemize}

In summary, there are 6 dynamical DOF (2 vector, 2 tensor and 2 scalar), all of which satisfy the 6D wave equation in the bulk. This is the correct number of propagating DOF in six-dimensional GR with azimuthal symmetry\footnote{Without this symmetry, there would be 3 additional dynamical DOF.}.

Below, we will also see that all of these 6 DOF can indeed be sourced.
This contradicts the claim in Ref.~\cite{Berkhahn:2012wg}, where only 5 sourced DOF were found, allowing to argue that the ghost mode ($ s $, see below) would not be dynamical. This wrong conclusion was reached by employing a gauge transformation which is in fact not allowed by the requirement of SO(2) symmetry, see Appendix~\ref{app:hamiltonian}.

\subsection{Junction conditions}

It remains to derive the linearized junction conditions~\eqref{eq:israel}. To this end, it is useful to perform a 3D tensor-vector-scalar decomposition of the (perturbation of the) energy-momentum tensor, analogous to~\eqref{eq:split_3D}:
\begin{subequations}
\begin{align}
	U_{0i} & = U_i^{(N)} + \partial_i U^{(L)} \,,\\
	U_{ij} & = U_{ij}^{(D)} + \partial_{(i} U_{j)}^{(V)} + \partial_i\partial_j U^{(B)} + \delta_{ij} U^{(S)} \,.
\end{align}
\end{subequations}
Energy conservation then decomposes into
\begin{subequations} \label{eq:en_cons}
\begin{align}
	- \dot U_{00} + \Delta_3 U^{(L)} & = - \frac{\lambda}{2\pi r_0} \dot \varphi \,,\\
	- \dot U^{(L)} + \Delta_3 U^{(B)} + U^{(S)}  & = - \frac{\lambda}{2\pi r_0} \varphi \,,\\
	- 2\dot U_i^{(N)} + \Delta_3 U_i^{(V)} & = 0 \,,
\end{align}
\end{subequations}
while $ U_{ij}^{(D)} $ and $ U_{\phi\phi} $ are unconstrained.

As discussed in more detail in Appendix~\ref{ap:junct_cond}, the junction conditions can be written as

\begin{itemize} %[leftmargin=*]

\item Tensor:
\begin{equation} \label{eq:israel_tensor}
	M_6^4 [D_{ij}'] + M_5^3 \Box_4 D_{ij}|_0 = U^{(D)}_{ij}
\end{equation}

\item Vector:
\begin{equation} \label{eq:israel_vector}
	M_6^4 [C_i']  + M_5^3 \Box_4 C_i|_0 = U^{(N)}_{i}
\end{equation}

\item Scalar:
\begin{subequations} \label{eq:israel_scalar}
\begin{equation}
%	M_6^4 \left ([C'] - \frac{\delta}{2\pi - \delta} \left (\varphi - s_0 \right) \right ) + M_5^3 \left.\left ( S + \varphi - Q \right)\right|_0 & = U^{(B)} \,,\\
	M_6^4 \left (  [J'] + \frac{\delta}{2\pi r_0} \varphi \right ) + M_5^3 \, \Box_4 J|_0 = - U^\mu_{\hphantom{\mu}\mu} + 3 U^{(S)} \,,\\
\end{equation}
\begin{multline}
	4 M_6^4 \left ( [s'] + \frac{\delta}{2\pi r_0} \varphi \right ) + M_5^3 \, \Box_4 \left ( s|_0 + 3 \varphi \right ) = \\
	- U^\mu_{\hphantom{\mu}\mu} + 3 \mixInd{U}{\phi}{\phi} \,. \label{eq:israel_s}
\end{multline}
\end{subequations}
\end{itemize}
These last two equations are not yet sufficient to solve for $ J $ and $ s $, because they contain $ \varphi $ as a third unknown on-brane function. It is determined by the $ (\phi\phi) $-junction condition, which can be rewritten as (see Appendix~\ref{ap:junct_cond})
\begin{equation} \label{eq:israel_phi}
	\frac{\delta}{2\pi - \delta} M_6^4 r_0 \Box_4 \left (\varphi - s|_0 \right ) + M_5^3 \Box_4 s|_0 = \mixInd{U}{\phi}{\phi} \,.
\end{equation}

As discussed in Sec.~\ref{sec:regularization}, we will now require the proper circumference of the regularized brane to be constant, implying $ \varphi = 0 $. (The case without stabilization is discussed in Appendix~\ref{ap:non_stabil}.) The appropriate $ \mixInd{U}{\phi}{\phi} $ which is needed to achieve this stabilization is determined by Eq.~\eqref{eq:israel_phi}, which now simplifies to
\begin{align} \label{eq:Uphiphi}
	\mixInd{U}{\phi}{\phi} = \left ( M_5^3 - \beta M_6^4 r_0 \right ) \Box_4 s|_0 \,, && \beta := \frac{\delta}{2\pi - \delta} \,.
\end{align}
The junction conditions for the two dynamical scalars then become
\begin{subequations} \label{eq:israel_scalar_stabil}
\begin{align}
	M_6^4 [J'] + M_5^3 \Box_4 J|_0 & = - U^\mu_{\hphantom{\mu}\mu} + 3 U^{(S)} \,, \label{eq:healthy_mode}\\
	4M_6^4 [s'] - 2 f M_5^3 \Box_4 s|_0 & = - U^\mu_{\hphantom{\mu}\mu} \,, \label{eq:israel_ghost}
\end{align}
\end{subequations}
where we defined the dimensionless constants
\begin{align}\label{eq:def_f}
	f := 1 - \frac{3\beta}{4\alpha} \,,  && \alpha := \frac{M_5^3}{2 r_0 M_6^4} \,.
\end{align}

All the junction conditions, viz.\ \eqref{eq:israel_tensor}, \eqref{eq:israel_vector} and \eqref{eq:israel_scalar_stabil}, now share the same, DGP-like structure. The only (but crucial) difference is that the BIG term in the junction condition for $ s $ comes with a negative sign if $ f>0 $. 

Note that $ f $ matches the function $ \hat f $ defined in Ref.~\cite{Niedermann:2014bqa} in the static case ($H=0$), where the sign of $ \hat f $ determined the stability of the model in the case of FLRW symmetries (at the nonlinear level).\footnote{Ref.~\cite{Niedermann:2014bqa} also uses a second regularization, where the extrinsic curvature in the interior is demanded to be static, implying a slightly different form of $\hat f$. We expect---although not explicitly studied here---to find the same accordance between both analyses in that case.} Therefore, one might already suspect that the scalar mode $ s $ will be a ghost in that parameter regime. We will now show on analytical grounds that this is indeed the case.

\subsection{Tachyonic ghost} \label{sec:tach_ghost}

We will use the vacuum to vacuum transition probability (in presence of an external source) as a diagnostic tool to probe for ghost modes. For the linear theory, it is given by
\begin{equation} \label{eq:vac_prob}
	\left | \langle 0 | 0 \rangle_T \right |^2 =  \exp \left [ - \operatorname{Im} (\mathcal{A}) \right ] \,,
\end{equation}
with
\begin{equation} \label{eq:def_A}
	\mathcal{A} := r_0 \int \!\rd^4 x\, \rd\phi \; \prescript{1}{}{T}^{(5)}_{\alpha\beta} h^{\alpha\beta}|_0 \,.
\end{equation}
Here, $ h^{\alpha\beta}|_0 $ should be evaluated at the classical solution in the presence of $ T_{\alpha\beta} $.
If the probability~\eqref{eq:vac_prob} is larger than one (or, equivalently, the imaginary part of $ \mathcal{A} $ is negative) then unitarity is violated\footnote{As is well known, this unitarity violation does not mean that the theory cannot be consistently quantized. Unitarity can indeed be restored by choosing a non-standard $ i\epsilon $ prescription for the ghost mode, reversing the sign of the ghost-residue in the propagator. However, what cannot be cured is the fact that the Hamiltonian is not bounded from below, which---as soon as interactions are included---causes a catastrophic instability. This instability is already present at the classical level, and has nothing to do with quantizing the theory. In any case, a ghost shows that the theory is pathological and thus useless.}, showing the existence of a ghost mode.

Since the ghost mode lies in the scalar sector, we can limit ourselves to a a source with vanishing tensor- and vector components.
Using the bulk equations to eliminate all constrained quantities, the source coupling term can---using integration by part and energy conservation~\eqref{eq:en_cons}---then be brought into the form
\begin{equation}
	\mathcal{A} = \frac{\pi\,r_0}{3}\int\!\rd^4 x\, \left [ 2 \left ( \mixInd{U}{\mu}{\mu} - 3U^{(S)} \right ) J|_0 +  \mixInd{U}{\mu}{\mu} s|_0 \right ] \,.
\end{equation}
Furthermore, it will be sufficient to consider a source satisfying $ \mixInd{U}{\mu}{\mu} = 3 U^{(S)} $, for which only $ s $ gets excited (i.e.\ $ J $ can consistently be set to zero) and the coupling term~\eqref{eq:def_A} simply reads
\begin{equation} \label{eq:A_ghost}
	\mathcal{A}^{(s)} = \frac{\pi\,r_0}{3} \int \!\rd^4 x \; \mixInd{U}{\mu}{\mu} s|_0 \,.
\end{equation}
In the following, it will be convenient to work in 4D Fourier space, i.e.\ we introduce
\begin{equation}
	\hat s(p, r) := \int \!\rd^4 x \; \re^{-\im p \cdot x} s(x, r) \qquad ( p \cdot x := p_\mu x^\mu )\,.
\end{equation}
The bulk equation~\eqref{eq:bulk_scalar_dyn} then becomes
\begin{equation} \label{eq:bulk_eom_FT}
	\left ( -p^2 + \Delta_2 \right ) \hat s = 0  \qquad ( p^2 := p_\mu p^\mu )\,,
\end{equation}
where the covariant 2D Laplace operator with respect to the deficit angle background geometry~\eqref{eq:met_bkg} reads (for $ \phi $-symmetric fields)
\begin{equation}
	\Delta_2 = \partial_r^2 + \frac{g'}{g} \partial_r =
	\begin{dcases}
		\partial_r^2 + \frac{1}{r} \partial_r & (r < r_0)\\
		\partial_r^2 + \frac{1}{r + \beta r_0} \partial_r & (r > r_0) \,.
	\end{dcases}
\end{equation}

The most general solution of~\eqref{eq:bulk_eom_FT}, which is continuous across the brane, regular at the origin, and falls off at radial infinity\footnote{{Furthermore, for $ p^2 < 0 $, i.e.~for modes which correspond to waves propagating in the bulk, one can check that the solution~\eqref{eq:s_bulk} corresponds to solely \emph{outgoing} radial waves, if the retarded prescription $ \operatorname{Im}(\omega) = +\epsilon $ is used, as would be appropriate for a classical calculation. This is an important consistency requirement as the brane is the only source of gravitational waves in the bulk. Note, however, that below we will use the Feynman prescription $ \operatorname{Im}(\omega^2) = +\epsilon $, since we are calculating the vacuum amplitude in the quantum theory.}}, is given by
\begin{equation} \label{eq:s_bulk}
	\hat s (p, r) = 
	\begin{dcases}
		\frac{I_0 ( r \sqrt{p^2})}{I_0( r_0 \sqrt{p^2})} \hat s|_0 & (r < r_0)\\
		\frac{K_0 ( \tilde r \sqrt{p^2})}{K_0( \tilde r_0 \sqrt{p^2} )} \hat s|_0 & (r > r_0) \,,
	\end{dcases}
\end{equation}
where $ I_n $ and $ K_n $ are the modified Bessel functions of the first and second kind, respectively, and $ \tilde r := r + \beta r_0 $. 
A priori, the solution is only defined for $p^2>0$. We find its analytic continuation by choosing the branch cut of the square root in the standard way, i.e.\ along the negative real axis.

%\footnote{\color{red}In the classical case, i.e. by employing the retarded $\omega$-integration contour, and for $p^2<0$ the solution \eqref{eq:s_bulk} corresponds to solely outgoing radial waves. This is an important consistency requirement as the brane is the only source of gravitational waves in the bulk.}

\begin{figure}
	\centering
		\includegraphics[width=0.45\textwidth]{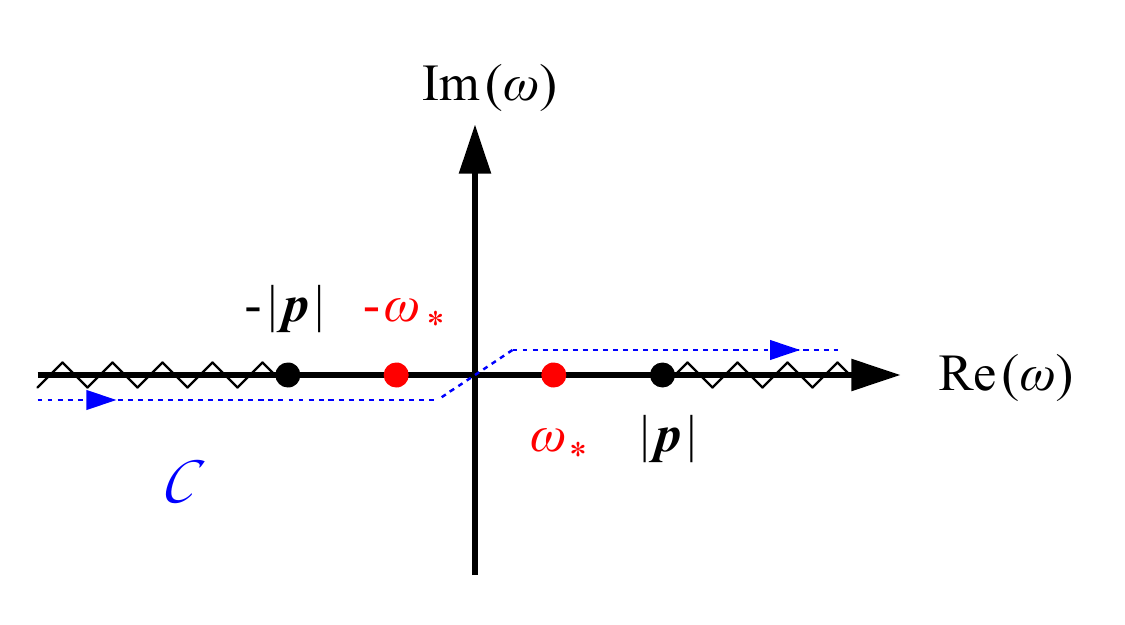}
		\caption{Analytic structure of the $ \omega \equiv p^0 $ dependence of the $ s $-propagator $ 1/Z(p) $, see Eq.~\eqref{eq:Z1}, for some fixed value $ \left |\vec{p} \right | \neq 0 $. The poles at $ \pm \omega_\ast $ correspond to the tachyonic ghost. For $ |\vec{p}| > m_\ast $ they lie on the real axis, between the origin and the branch cuts starting at the poles at $ \pm \left |\vec{p}\right | $; for $ |\vec{p}| < m_\ast $ they lie on the imaginary axis. The contour $\mathcal{C}$ (dotted blue line) indicates the Feynman-contour of integration.}
		\label{fig:complex_plane}
\end{figure}

%In particular, for $p^2<0$, and taking into account the Feynman prescription, $\operatorname{Im} \left(p^2\right)=- \epsilon$, we find at radial infinity
%
%\begin{align}
%	K_0 ( \tilde r \sqrt{p^2}) \propto \re^{\im r \sqrt{-p^2}} && (r \to \infty)\;,
%\end{align}
%
%which corresponds to an outgoing radial wave. This is an important consistency requirement, as the brane is the only source of gravitational waves in the bulk. Moreover, it excludes the possibility that the ghost instability is an artifact caused by any incoming waves. 
Plugging \eqref{eq:s_bulk} into the junction condition~\eqref{eq:israel_ghost} yields
\begin{equation}
	\frac{4M_6^4}{r_0} Z(p)\, \hat s |_0 = - \hat U^\mu_{\hphantom{\mu}\mu} \,,
\end{equation}
with the inverse $ s $-propagator
\begin{equation}
	\label{eq:Z1}
	Z(p) := \alpha f z^2 - \,z \left ( \frac{I_1( z)}{I_0(z)} + \frac{K_1( (1 + \beta) z)}{K_0( (1 + \beta) z)}  \right ) \,,
\end{equation}
where we introduced the dimensionless variable 
\begin{align}
	  z 	&:= \, r_0 \sqrt{p^2} \nonumber \\
 	 	&\phantom{:}\equiv r_0 \sqrt{ \vec{p}^2-\omega^2}\,.
\end{align}
The source coupling term~\eqref{eq:A_ghost} finally becomes
\begin{equation} \label{eq:s_vertex}
	\mathcal{A}^{(s)} = - \frac{r_0^2}{24M_6^4} \int \frac{\rd^4 p}{(2\pi)^3} \left |\mixInd{\hat U}{\mu}{\mu} (p) \right |^2 \frac{1}{Z(p)} \,,
\end{equation}	
%
%To investigate the analytic structure of the integrand, it is necessary to specify the branch cuts that appear due to the square-roots and Hankel functions. To this end, it is useful to first rewrite the inverse propagator as
where the branch cuts and poles in the $\omega$-integration are surrounded according to the Feynman prescription, i.e.\ $\operatorname{Im}  \left( \omega^2 \right) = +\, \epsilon$ along the integration contour.
%, and
%
%\begin{equation}
%	w := -\im z = r_0 \sqrt{p^2} \equiv r_0 \sqrt{\vec{p}^2 - \omega^2} \,.
%\end{equation}
%
%We will now choose the branch cuts of the square-root and the Bessel $ K $ functions in the standard way, i.e.\ running from the origin along the negative real axis. 
The analytic structure of $ Z^{-1} $, for some fixed value $ |\vec{p}| \neq 0 $ (and the case $ f > 0 $) is shown in Fig.~\ref{fig:complex_plane}. The branch cuts along the real axis can be interpreted as a continuum of gap-less Kaluza Klein modes, like in the DGP model~\cite{Dvali:2007kt}. They are also present in pure 6D GR with a cylindrical source, and are thus not expected to cause any problems. Below, we will confirm that this is indeed the case.

\begin{figure}
	\centering
		\includegraphics[width=0.45\textwidth]{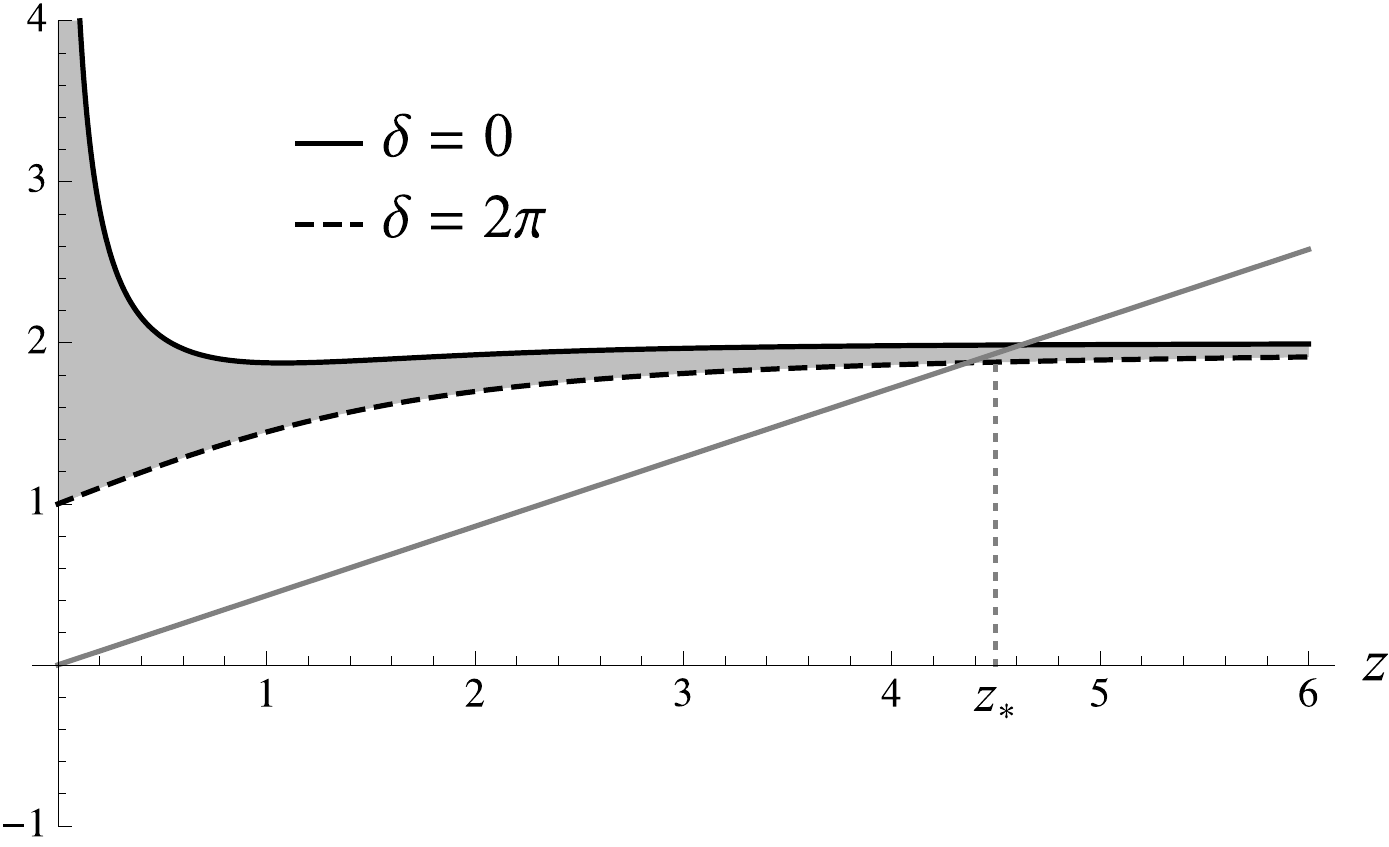}
		\caption{Graph of the right-hand side of Eq.~\eqref{eq:ghost_mass}, which determines the ghost mass $ m_\ast \equiv z_\ast / r_0 $. For values of the deficit angle $ \delta $ between $ 0 $ and $ 2\pi $, the curve lies in the shaded region. It always goes to $ 2 $ as $ z \to \infty $. The gray line corresponds to the left-hand side of the equation, for some positive value of $ f $. For $ f < 0 $ its slope is negative and there is no solution $ z_* $.}
		\label{fig:rhs}
\end{figure}

%We also explicitly checked (numerically) that---at least for an $ \omega $-independent source---their contribution to the imaginary part of $ \mathcal{A}^{(s)} $ is positive, see Fig.~\ref{fig:branchCut}.

However, for $ f>0 $, there are additional isolated poles at
\begin{equation} \label{eq:ghost_disp}
	\omega = \pm \sqrt{\vec{p}^2 - m_\ast^2} =: \pm\, \omega_\ast \,,
\end{equation}
where $ m_\ast $ is given by $ m_\ast = z_\ast / r_0 > 0 $, with $ z_\ast $ being the solution of
\begin{equation} \label{eq:ghost_mass}
	\alpha f z_\ast = \frac{I_1(z_\ast)}{I_0(z_\ast)} + \frac{K_1((1 + \beta) z_\ast)}{K_0((1 + \beta) z_\ast)} \,.
\end{equation}
The right hand side of this equation is plotted in Fig.~\ref{fig:rhs}, and shows that there is indeed a solution $ z_\ast > 0 $ if and only if $ f>0 $. 
The negative sign of the mass term in the dispersion relation~\eqref{eq:ghost_disp} shows that this pole in the propagator of the scalar mode $ s $ is a tachyon, implying the existence of exponentially growing solutions for $ s|_0(t) $. Below, we will show that it is also a ghost, in agreement with the result in Refs.~\cite{Dubovsky:2002jm, Hassan:2010ys}, but generalizing it to a background with nonzero deficit angle.

Even though~\eqref{eq:ghost_mass} cannot be solved analytically, one can obtain the asymptotic formula for $ m_\ast $ in the physically relevant limit $ \alpha \to \infty $ (i.e.~$ r_0 \ll M_5^3 / M_6^4 $) by expanding the Bessel function for small arguments, yielding
\begin{equation} \label{eq:m_ghost_asympt}
	m_\ast^2 \sim  \frac{1 - \delta / (2\pi)}{r_0^2 f \alpha \ln\alpha} \qquad (\alpha \to \infty) \,.
\end{equation}
Note that for $ \delta = 0 $ (and neglecting the small logarithmic correction) this agrees with the tachyon mass derived in~\cite{Dubovsky:2002jm, Hassan:2010ys}, viz.\ $ m_\ast \sim M_6^2 / M_4 $. However, the nontrivial deficit angle background leads to an important modification: as $ \delta $ increases, $ f $ approaches zero and Fig.~\ref{fig:rhs} shows that the tachyon then becomes infinitely heavy, as the intersection moves to larger values of $z$. When the threshold $ f = 0 $ is crossed, the pole finally disappears completely.

\begin{figure}
	\centering
		\includegraphics[width=0.45\textwidth]{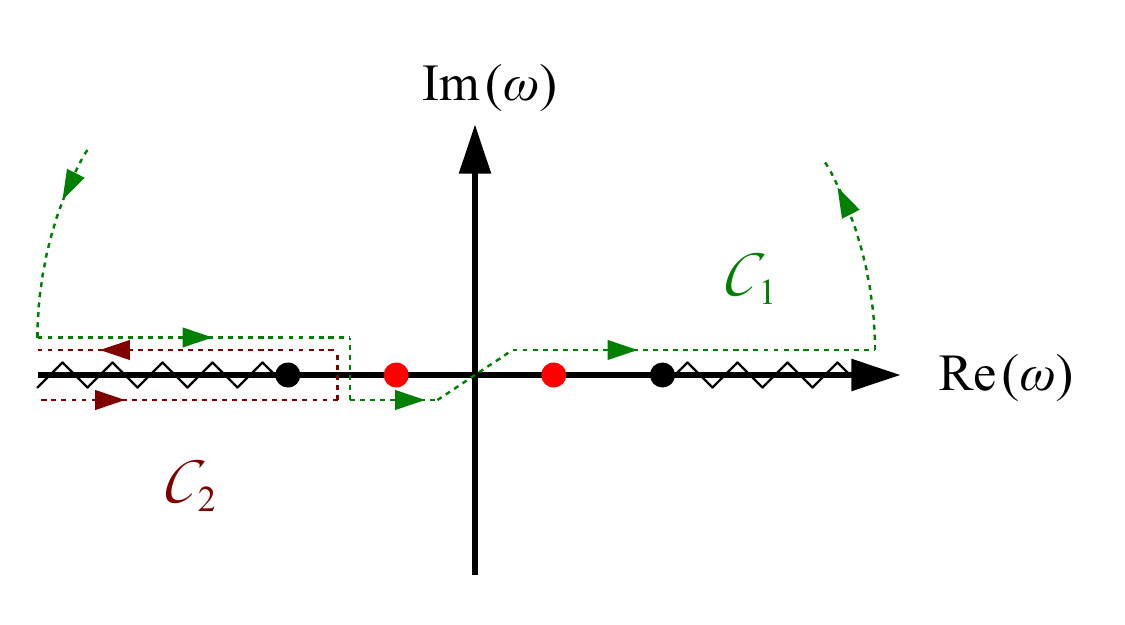}
		\caption{Decomposition of the Feynman-contour into a closed path around one of the poles ($\mathcal{C}_1$) and a branch cut contribution ($\mathcal{C}_2$). The half circle, which closes the contour at infinity, does not contribute to $ \mathcal{A}^{(s)} $.}
		\label{fig:complex_plane_2}
\end{figure}

To disentangle the tachyon and branch cut contribution to  $\mathcal{A}^{(s)}$, we consider two independent integration contours in the complex $\omega$-plane: a closed path $\mathcal{C}_1$ encircling one of the poles, and another open path $\mathcal{C}_2$ running along both sides of the branch cut in opposite directions, see Fig.~\ref{fig:complex_plane_2}. It can be shown that the half circle in $\mathcal{C}_1$ does not contribute to the amplitude\footnote{In Fig.~\ref{fig:complex_plane_2} we only show the case when the contour has to be closed in the upper half-plane, in which the pole and branch cut on the negative real axis contribute. But one can easily check that the other case yields exactly the same result.}. It thus follows that the original integration along $\mathcal{C}$ can be decomposed into the sum of the two contributions $\mathcal{C}_1$ and $\mathcal{C}_2$.

\begin{figure}
	\centering
		\includegraphics[width=0.45\textwidth]{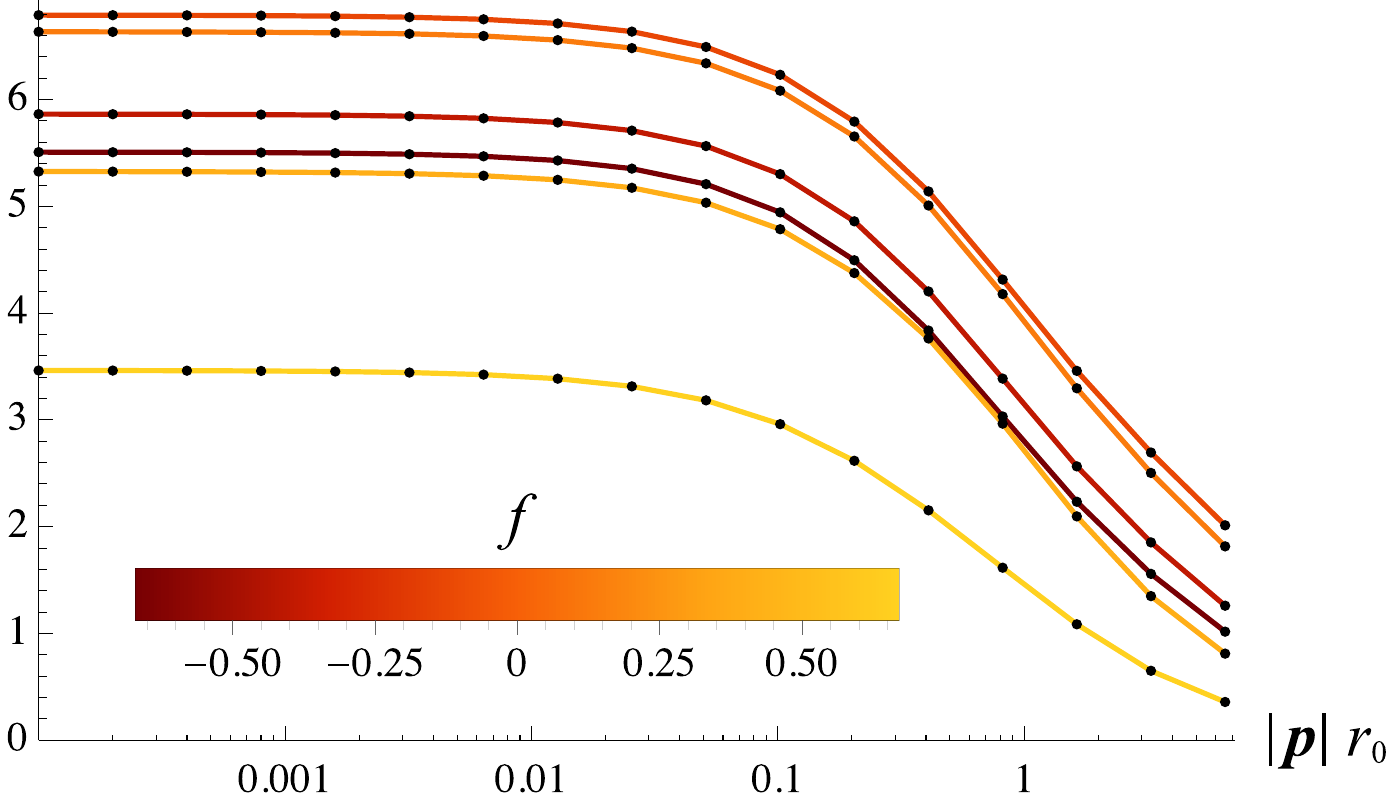}
		\caption{Numerical evaluation of the branch-cut contribution, viz.~$ - \operatorname{Im} \left [ r_0 \int_{\mathcal{C}_2} \!\rd\omega \, Z^{-1}(\omega, \vec{p}) \right ] $. The positive values imply that there are no ghost modes, irrespective of the sign of $ f $. Here, the deficit angle was chosen as $ \delta = \pi $, but other values do not change this result.}
		\label{fig:branchCut}
\end{figure}

As for the branch cut contour $\mathcal{C}_2$, we checked numerically that---at least for an $\omega$-independent source---its contribution to the imaginary part of $\mathcal{A}^{(s)}$ is positive, see Fig.~\ref{fig:branchCut}. Hence, the branch cut contains no ghost modes.

\begin{figure}[hbt]
	\centering
		\includegraphics[width=0.45\textwidth]{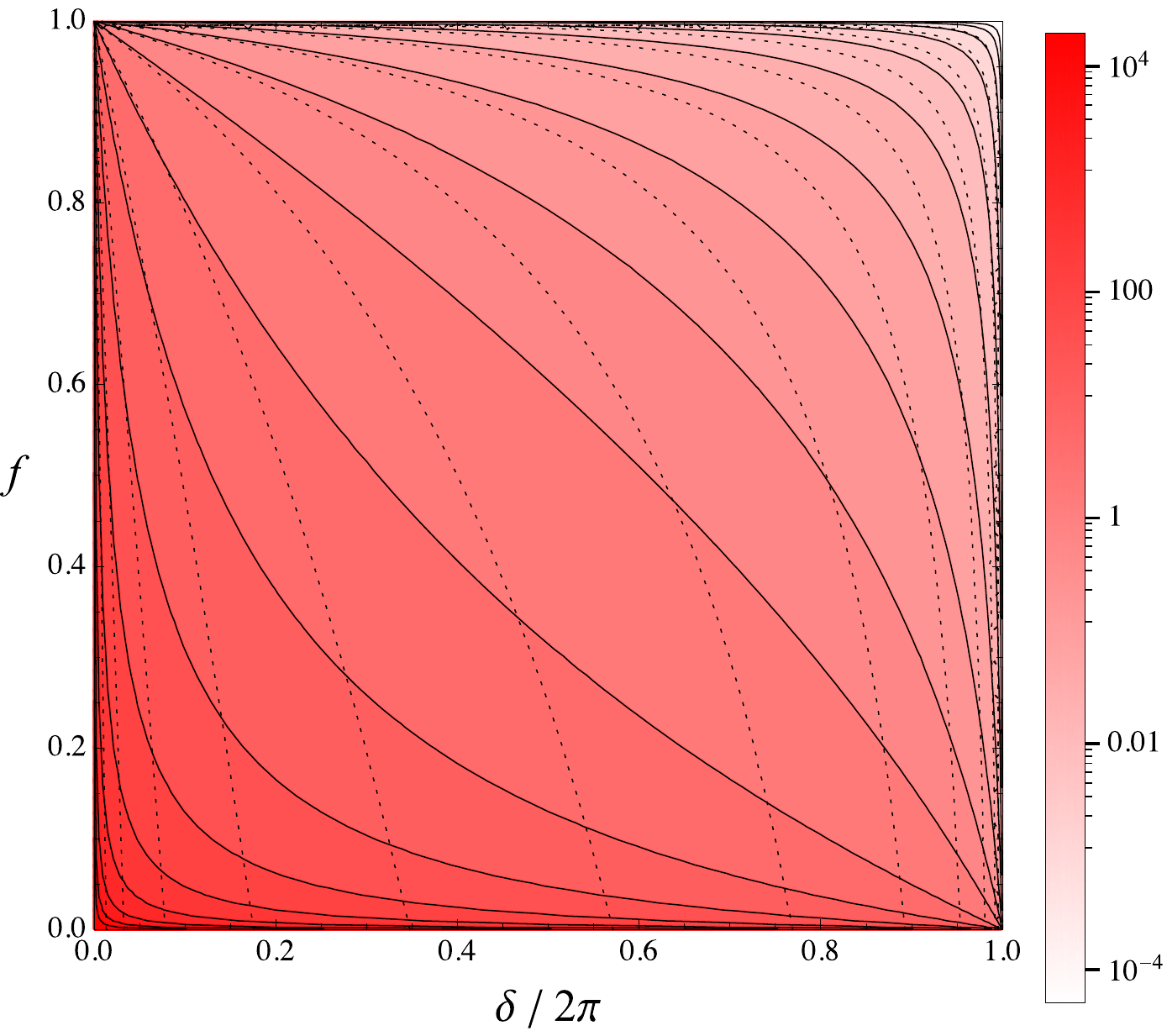}
		\caption{Contour plot of the ghost residue~\eqref{eq:ghost_res} times $ \mp\omega_\ast r_0^2 $, as a function of the deficit angle $ \delta $ and $ f $ as defined in~\eqref{eq:def_f}, showing that it is indeed positive in the whole parameter space (in which the pole at $ \omega_\ast $ exists, i.e.\ for $ f>0 $). The dotted lines are lines of constant $ \alpha  $, which $ \to 0 $ on the left and $ \to \infty $ on the right.}
		\label{fig:residue}
\end{figure}

The contribution of $\mathcal{C}_1$ to the amplitude is proportional to the sum of the residues of all enclosed poles. Therefore, to show that the tachyon---in the parameter region where it exists---is also a ghost, let us investigate the residue of this pole for the case $ \left |\vec{p}\right | > m_\ast $, i.e.\ when the poles lie on the real axis (as in Fig.~\ref{fig:complex_plane}). (For momenta $ |\vec{p}| < m_* $, the pole lies on the imaginary axis and only contributes to the real part of $ \mathcal{A}^{(s)} $, which does not affect the vacuum transition probability~\eqref{eq:vac_prob}. Physically speaking, the ghost can only be excited for momenta larger than its mass.) A straightforward calculation gives
\begin{multline} \label{eq:ghost_res}
	\mathrm{Res} \left (\frac{1}{Z(p)}, \,\omega = \pm\, \omega_\ast \right ) = \mp \frac{1}{\omega_\ast r_0^2} \Biggl [ 2 \alpha f + \beta  \\
	+ \left ( \frac{I_1}{I_0} \right )^2 - (1+\beta) \left (\frac{K_1}{K_0} \right )^2 \Biggr ]^{-1} \,.
\end{multline}
Here, the arguments of the Bessel $ I $ and $ K $ functions are $ z_\ast $ and $ (1+\beta) z_\ast $, respectively.
It turns out that the expression in square brackets, when evaluated numerically, is always positive. However, we did not succeed in extracting this information analytically, and therefore Fig.~\ref{fig:residue} shows the contour plot of the residue---leaving out the overall factor $ \mp 1 / (\omega_\ast r_0^2) $---as a function of the two independent model parameters $ \delta / (2\pi) \equiv \beta / (1+\beta) $ and $ f \equiv 1 - 3 \beta / (4\alpha) $. For a non-negative, sub-critical tension we have $ \delta \in [0, 2\pi) $ and $ f \leq 1 $. Furthermore, as already discussed, the ghost pole only exists for $ f > 0 $. Thus, the plot in Fig.~\ref{fig:residue} covers the whole relevant parameter space, and one can see that the expression in square brackets is indeed always positive.

One can easily check that this result leads to a negative\footnote{The sign from the overall factor in \eqref{eq:ghost_res} is compensated by the one from the negative/positive orientation when encircling the pole at $ \pm \omega_\ast $, cf.\ Fig.~\ref{fig:complex_plane_2}. Then, there is one more minus sign from the explicit overall factor in~\eqref{eq:s_vertex}.} imaginary part of $ \mathcal{A}^{(s)} $, corresponding to a ghost. This ghost mode can be excited for 3-momenta $ \left |\vec{p}\right | $ larger than the ghost mass $ m_\ast $. However, since the ghost is also a tachyon, which can be excited with arbitrarily low momenta, the linearized theory is completely unstable (for all momenta), if $ f > 0 $.

\begin{figure}[htb]
	\centering
		\includegraphics[width=0.45\textwidth]{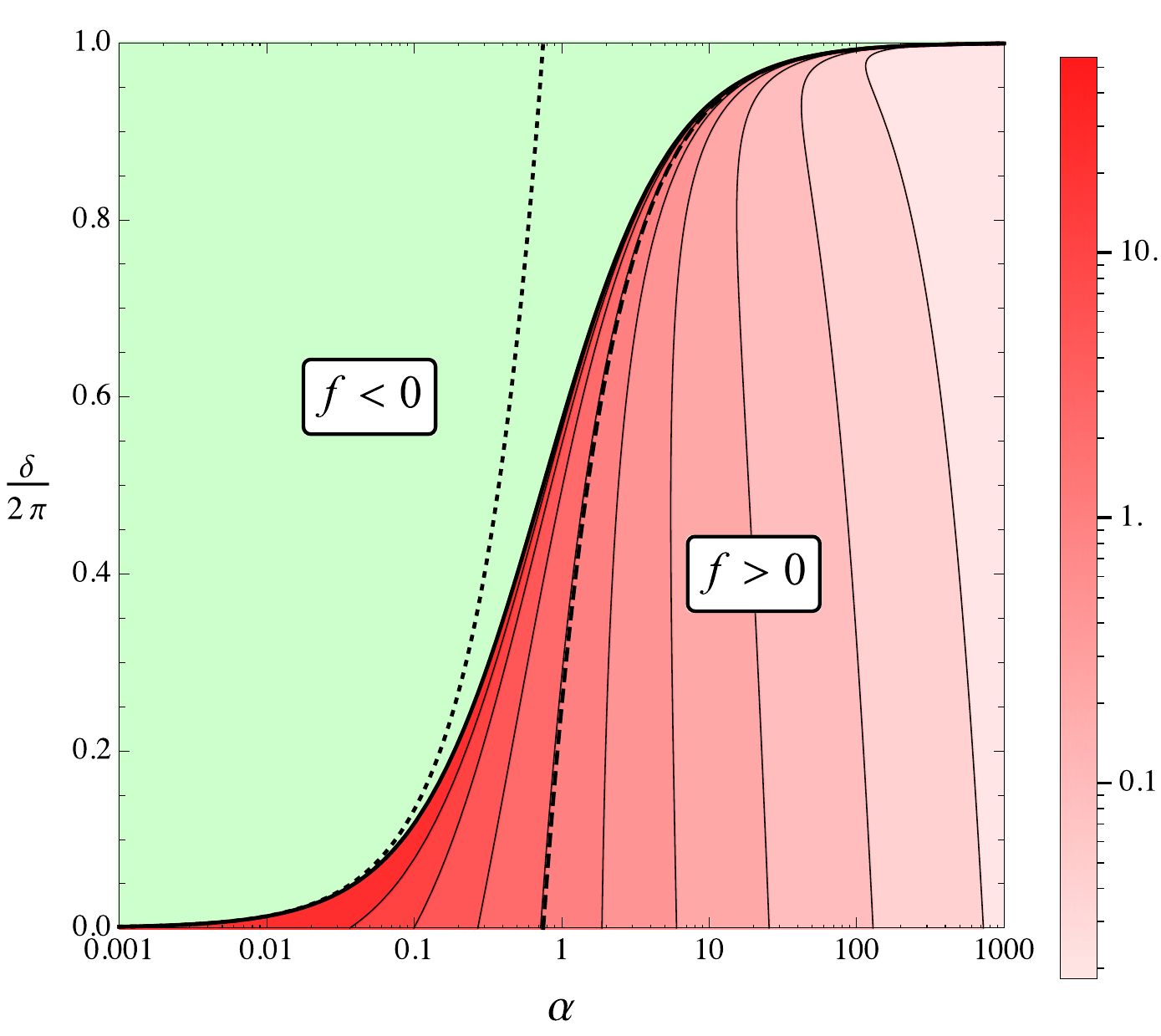}
		\caption{Stability of the linearized theory is determined by the two model parameters $ \alpha \equiv M_5^3 / (2 M_6^4 r_0) $ and deficit angle $ \delta $. The tachyonic ghost only exists in the red region ($ f>0 $), where the contours indicate its mass (in units of $ 1/r_0 $). The dashed line corresponds to the stability bound in the alternative regularization of~\cite{Niedermann:2014bqa}. The dotted line shows the (somewhat stronger) bound on the model parameters discussed below Eq.~\eqref{eq:f_EFT}.}
		\label{fig:cont}
\end{figure}

If, on the other hand, $ f < 0 $, the pole and thus the tachyonic ghost is absent and the model is stable. The stable and unstable regimes are visualized in a parameter plot in Fig.~\ref{fig:cont}. It also shows that the tachyon mass diverges as the borderline $ f=0 $ is approached. Thus, the tachyonic instability is more severe close to the stability bound. 

\subsection{Regularization independence}

In Sec.~\ref{sec:regularization} we argued that we are insensitive to the details of the regularization as long as we pose low energy questions. In the context of the above analysis this implies that only ghost poles with a mass $ m_\ast r_0 \ll 1  $ are a generic prediction in the sense that they are expected for other types of UV models.  Contrariwise, this means that the precise shape of the stability line in Fig.~\ref{fig:cont} depends on the way of regularizing---with the uncertainty region corresponding to the dark shaded red region approximately left of the $\alpha=1$ line. 

This UV dependence was carefully discussed in~\cite{Niedermann:2014bqa} where two different regularizations were used: 
\begin{itemize}

\item
 One regularization fully resolves the gravitational sector inside and outside the ring, hence matching exactly the regularization used here. A generalized version\footnote{It is denoted by $\hat f(\tau)$ and discussed in the appendix of Ref.~\cite{Niedermann:2014bqa}. } of $f$ valid for FLRW symmetries was derived, which explicitly depends on the Hubble parameter $H$. Since $H$ is zero for the background solution \eqref{eq:def_angle_sol}, it would enter in \eqref{eq:def_f} only through higher order corrections. The important point is that by setting $H$ to zero, we exactly reproduce the result here. In other words, the old analysis is in full agreement with the parameter plot shown in Fig.~\ref{fig:cont}.

\item
Another regularization neglected the interior dynamics.  Consequently, the generalized version of $f$ was slightly different, in particular, there the borderline $ f=0 $ crossed the $ \delta=0 $ axis, leaving a small stable regime near $ \alpha = 0 $ in the tensionless case, unlike in the other regularization. The corresponding modified stability bound is shown as the dashed line in Fig.~\ref{fig:cont}. However, it can be checked that the difference is limited to the regime with $ m_\ast r_0 \gtrsim 1  $, as expected. In particular, this does not affect the phenomenologically interesting regime characterized by $ \alpha \gg 1 $, as discussed in Sec.~\ref{sec:pheno}.

\end{itemize}

In summary, we could show that the regularization dependence found in~\cite{Niedermann:2014bqa} is bound to a regime where the inner structure of the brane is already being resolved. This had to be expected from an EFT perspective.

%These results are in perfect agreement with the non-linear analysis in~\cite{Niedermann:2014bqa}, as will be discussed in more detail in Sec.~\ref{sec:pheno}.

\section{Hamiltonian analysis}
\label{sec:hamiltonian}

Above, we have shown that a tensionless brane with large enough BIG scale is necessarily plagued by a scalar ghost mode. In this section, we gather strong evidence that the same result holds true in higher dimensions ($D>6$). Thereby, we also provide a complementary picture to support the previous result in six dimensions. To be concrete, we show that the ghost mode leads to a \emph{negative} contribution to the energy density at the brane position. As a diagnostic tool we use the Hamiltonian on the constraint surface that we derive for the sourceless theory, i.e.\ $U_{\alpha\beta}=0$. This technique was also used in an earlier work \cite{Berkhahn:2012wg}. However, there it led to the erroneous claim that the theory in 6D without a brane-tension is ghost-free. The reason for that failure is discussed in Appendix~\ref{app:hamiltonian}. The purpose here is to present a corrected analysis which also extends to higher dimensions.

We will study small metric perturbations on a bulk (and brane) Minkowski background, $g_{AB}=\eta_{AB}+h_{AB}$. To that end, we use a decomposition of $h_{AB}$ that generalizes the one used in \eqref{eq:split_3D_0}, \eqref{eq:split_ri} and \eqref{eq:def_b_s} to arbitrary dimensions. To be precise, we introduce
\begin{subequations}
\begin{align}
	h_{0a} & = n_a + \partial_a l \,, \\
	h_{ab} & = d_{ab} + \partial_{(a} v_{b)} + \partial_a \partial_b b + \delta_{ab} s \,, \label{eq:split_ab}\\
	h_{ia} & = E_{ib} + \partial_{i} F_a + \partial_a G_i + \partial_i \partial_a H \,,
\end{align}
\end{subequations}
where under the SO($n$) group $d_{ab}$ transforms as a transverse and traceless tensor, $E_{ib}$, $v_a$, $F_a$, $n_a$ as transverse vectors and $G_i$ as a scalar. Moreover, $G_i$ and $E_{ia}$ are 3D vectors, while all the other fields defined above are 3D scalars. Note that the definitions for the 3D spatial components in \eqref{eq:split_0i} and \eqref{eq:split_ij} still apply.

Instead of fixing a particular gauge, we will again work with gauge invariant variables, i.e.\ combinations of the fields invariant under \eqref{eq:gauge}. To that end, we extend the definitions in \eqref{eq:g_inv_var_3D} by
\begin{subequations}
\begin{align}
	J & := 3S + (n-1) s &&\\
		& E_{ib}\,, 				&		& d_{ab}\,, \\
	w_a &:= 2 F_a - v_a\,,  			& c_a 	&:= n_a - \dot F_a\,.
\end{align}
\end{subequations}%

A crucial benefit of the constraint analysis is that it corresponds to a simple reformulation of the theory and does not require to find any explicit solution. As a consequence, we do not have to regularize the setup to avoid singularities and can instead work with an infinitely thin defect described by $n$-dimensional $\delta$-functions. We will check the validity of this approach a posteriori by comparing the physical degrees of freedom and their matching equations in the special case of $n=2$ to the results derived in Sec.~\ref{sec:lin_stab} for a (stabilized) brane when the tension is set to zero. We will find perfect agreement. 

Let us stress that this is also a nice way of demonstrating that our results---in particular statements about the dynamical content of the theory---do not depend on our way of regularizing the brane as a ring in extra space. However, we will see that the two descriptions are only equivalent if the (proper) circumference of the ring is demanded to be constant. This is clear since without such a  stabilization the regularization introduces an additional degree of freedom corresponding to the angular size modulus of the brane, which is of course not present for an infinitely thin defect. In our notation it is given by the field $ \varphi $. As discussed in Sec.~\ref{sec:regularization}, whether or not we should include such a mode depends on the UV model. While in the main part of the paper (including this section) it is set to zero, we discuss its effects in Appendix~\ref{ap:non_stabil}. 

In this section we will use Cartesian coordinates $X^A=(t, x^i, y^a)$, such that the brane is at the constant coordinate position $y^a=0$. The d'Alembert and Laplace operator is denoted as before but generalized to $D$ dimensions or $n$ codimensions, viz.\ $\Box_D$ or $\Delta_n$.
\subsection{Reduced Lagrangian}
Due to gauge invariance, it is possible to fully express the Lagrangian $\mathcal{L}$ in terms of gauge invariant variables. The calculation is straightforward but little enlightening; thus, we do not display the rather lengthy resulting expression. 
Varying it with respect to the Lagrange multipliers $P$ and $Q$ yields the constraint
\begin{equation}\label{eq:constr_P}
	2 \dot J + \Delta_3 \left ( \dot O + P \right )  = 0\,,
\end{equation}
which is identical to \eqref{eq_bulk_scalar_P} and allows to eliminate $P$,  as well as
\begin{multline}\label{eq:constr_C}
	M_D^{D-2} \Big[ (2\Delta_{3}+3\Delta_{n})J + (2+n)\Delta_{3}s + 3\Delta_{3}\Delta_{n}{O} \Big] \\
	+ M_4^2\, \delta_n(y) 2 \Delta_{3} \Bigl [J - (n-1)s \Bigr] = 0\,,
\end{multline}
constraining $ O $.

Before we continue, let us check that the last equation is indeed a generalization of what we found before.
Away from the brane, we recover \eqref{eq_bulk_scalar_c} by setting $n=2$. Due to the $\delta$-term, this equation also contains specific on-brane information, which can be compared to the matching equations derived in the last section. This is easily achieved by regularizing as described in Sec.~\ref{sec:regularization}, which amounts to the replacement
\begin{align}
	\delta_2(y) \rightarrow \frac{\delta(r-r_0)}{2\pi r_0}\;.
\end{align}
Integrating the resulting equation over the interval $[r_0-\epsilon, r_0+\epsilon]$ and identifying $2\pi r_0 M_5^3 \equiv M_4^2$, one indeed recovers the $ (00) $ junction condition\footnote{There, the $ (00) $-junction condition is a certain linear combination of Eqs.~\eqref{eq:healthy_mode}, \eqref{eq:israel_ghost} and \eqref{eq:jump_c} with $\delta=0$ and $ \varphi=0 $.} from Sec.~\ref{sec:lin_stab}.
This agreement can similarly be checked for all remaining equations of motion.

A third scalar constraint for $Q$ arises in this language by differentiating \eqref{eq:constr_P} with respect to time and using another dynamical equation (which follows from varying the action with respect to $O$) to eliminate $\ddot O$. We find\footnote{In the terminology of the Dirac constraint formalism, this would correspond to a tertiary constraint (while \eqref{eq:constr_P} and \eqref{eq:constr_C} are secondary constraints).}
\begin{align}
	3 \Delta_3 Q - 2 \Delta_3 J + 3\ddot J -(n-1)\Delta_3 s =0\,.
\end{align}
A similar calculations allows to recover two vector constraints, one identical to~\eqref{eq:bulk_vector_constr}, and another one constraining $w_a$; for the sake of completeness, 
\begin{align}
	\Delta_3 W_i + 2 \dot C_i & = 0\,,\\
	\Delta_n w_a + 2 \dot c_a & = 0\,.
\end{align}

We now use the constraints to eliminate the $(4+n)$ non-dynamical variables $O$, $P$, $Q$, $W_i$ and $w_a$ (as well as their time derivatives) in the Lagrangian, which consequently is expressed solely in terms of dynamical degrees of freedom. The resulting Lagrangian is diagonal, and by decomposing it into its tensor, vector, and scalar contributions, $\mathcal{L}=\mathcal{L}_{\rm T} +\mathcal{L}_{\rm V} +\mathcal{L}_{\rm S} $, we get:

\begin{itemize} 

\item Tensor:
\begin{multline}
	4\,\mathcal{L}_{\rm T} = 	 
		M_D^{D-2} 
		\Big[
		- \left(\partial_A D_{ij}\right)^2 
  		-\left(\partial_A d_{ab}\right)^2 
		\Big]\\
		\!+ M_4^2  
		\Big[
		- \left(\partial_\mu D_{ij} \right)^2
		\Big] 	\delta_n(y)
		\,.						
\end{multline}
%
%where $\left(\partial_A f_{ij}\right)^2= \partial_A f_{ij}\partial^A f^{ij} $.
The 3D tensor describes two DOF which also have kinetic support on the brane. They correspond to the helicity-two modes in 4D GR and are thus crucial in realizing a 4D regime. The extra space tensor carries $ (n+1) (n-2) / 2 $ DOF which all decouple from the brane. In particular, they would not couple to a localized source on the brane.
\item Vector:
\begin{multline}
	2\,\mathcal{L}_{\rm V} = 	 
		M_D^{D-2} 
		\Big[
		- \left(\partial_A C_{i}\right)^2 
  		-\left(\partial_A c_{a}\right)^2 
		-\left(\partial_A E_{ia}\right)^2 
		\Big]\\
		\!+ M_4^2  
		\Big[
		- \left(\partial_\mu C_{i} \right)^2
		\Big] 	\delta_n(y)
		\,.						
\end{multline}
There are two vector DOF, described by $C_i$, which have a standard DGP-type action and hence a localized kinetic term. The remaining $3(n-1)$ vector DOF are decoupled from the brane.
\item Scalar:
\begin{multline}
	6 \, \mathcal{L}_{\rm S}=
		 M_D^{D-2} 
		\Big[
		- (\partial_A J)^2 
		- \frac{(n-1)(n+2)}{2} (\partial_A s)^2
		\Big]\\
		+M_4^2  
		\Big[
		- (\partial_\mu J)^2 +  (n-1)^2 (\partial_\mu s)^2 
		\Big] \delta_n(y)
\end{multline}
We find two DOF in the scalar sector. They are decoupled from each other, and, as expected, $s$ has a wrong sign kinetic term on the brane. Also note that for $n=1$ the scalar $s$ simply disappears, confirming that the codimension-one model is indeed ghost-free on a Minkowski background.
\end{itemize}
In summary, for $n>1$ there are always six degrees of freedom that can be excited by an on-brane source. Five of them ($D_{ij}$, $X_i$ and $J$) have a standard DGP-type action, whereas the scalar $s$ comes with a wrong sign kinetic term on the brane---irrespective of the number of codimensions. Therefore, tuning the tension to zero seems to cause the same ghost pathology in any higher dimension. In addition, there are further $[n(n+5)/2-4]$ DOF that only propagate in the bulk and are invisible to a brane observer. In total, we have $D(D-3)/2$ DOF corresponding exactly to the number of propagating degrees of freedom in $D$-dimensional Einstein gravity. This result strongly supports the EFT picture, according to which the induced gravity term arises simply by integrating out heavy particles in the presence of higher dimensional GR. From that perspective it is not surprising that we find the same number of propagating degrees of freedom.
In the codimension-two analysis in Sec.~\ref{sec:lin_stab} we assumed the fluctuation to respect SO(2) symmetry, which effectively eliminated three DOF\footnote{A similar assumption would here eliminate all degrees of freedom that do not couple to on-brane sources.}. Since the present analysis does not rely on that assumption, the result also holds in the case of additional sources in the bulk, which, in general, would spoil the SO($n$) symmetry. 

Let us stress that our calculation is also valid for one codimension; in that special case $s$ completely drops out of the equations, thus leaving a theory of five healthy degrees of freedom. 

\subsection{Reduced Hamiltonian}
A ghost mode necessarily leads to a classical instability, which manifests itself in a Hamiltonian that is \textit{not} bounded from below. In the remainder of this section, we will show that excitations of the ghost mode indeed lower the local energy density at the brane, thereby showing a pathology\footnote{Note that we did not calculate the vacuum persistence amplitude for $n>2$, but we also expect a violation of unitarity.}. Since all tensor and vector modes have the same DGP-type action as the scalar $J$, it suffices to investigate the scalar sector only. The only difficulty in deriving the Hamiltonian consists in a proper treatment of the localized terms. This can be consistently done by decomposing the conjugate momentum fields
\begin{align}
	\tilde\Pi_J := \delta\mathcal{L}/\delta \dot J && \text{and} && \tilde\Pi_s := \delta\mathcal{L}/\delta \dot s
\end{align}
into a regular and an irregular part according to  
\begin{subequations}
\begin{align}
3\,\tilde\Pi_J &=  \Bigl[ M_D^{D-2} + M_4^2\, \delta_n(y) \Bigr] \Pi_J \,,\\
6\,\tilde\Pi_s		&=  \Bigl[ (n-1)(n+2)\, M_D^{D-2}\, \nonumber\\
			&\qquad\qquad\qquad
			-2(n-1)^2\,M_4^2\, \delta_n(y) \Bigr] \Pi_s\,,
\end{align}
\end{subequations}
where we defined
\begin{align}
	\Pi_J := \dot{J} && \text{and} && \Pi_s := \dot{s} \,.
\end{align}
This decomposition is well defined because $\dot J$ and $\dot s$ are regular functions at the brane (only $ r $-derivatives would introduce discontinuities). The scalar Hamiltonian density for both fields,
\begin{align}
	\mathcal{H}_{J} := \tilde\Pi_J \dot J - \mathcal{L}_{J} && \text{and} && \mathcal{H}_{s} :=  \tilde\Pi_s \dot s - \mathcal{L}_{s} \,, 
\end{align}
can then be readily derived. For the healthy scalar we get a contribution
\begin{multline}
	6\, \mathcal{H}_{J}= M_D^{D-2} 
	\Bigl[ 
		\left(\Pi_J \right)^2+\left(\partial_i J \right)^2+\left(\partial_a J \right)^2
	\Bigr]\\
	+ M_4^2 \Bigl[\left(\Pi_J \right)^2 + \left(\partial_i J \right)^2
	\Bigr]\delta_n(y)\,,
\end{multline}
which indeed shows that $J$ contributes positively to the energy of the system. The tensor- and vector contributions take the same (manifestly positive) form with $J$ replaced by $D_{ij}$ and $C_i$, respectively. Hence, they also constitute healthy (brane-coupled) fields. However, the expression for the scalar $s$ becomes 
\begin{multline}
	12\, \mathcal{H}_{s}= (n-1)(n+2)\, M_D^{D-2} 
	\Bigl[ 
		\left(\Pi_s \right)^2+\left(\partial_i s \right)^2+\left(\partial_a s \right)^2
	\Bigr]\\
	-2(n-1)^2\, M_4^2\Bigl[\left(\Pi_s \right)^2+\left(\partial_i s \right)^2
	\Bigr]\delta_n(y)\,,
\end{multline}
thus displaying a \emph{negative} energy contribution at the brane position. In accordance with the result of the last section, it is present for an arbitrarily small coefficient $M_4$ and only vanishes if the induced term is set to zero exactly. In the latter case our result simply reflects the stability of higher dimensional Einstein gravity.

As an aside, note that we also derived the Hamiltonian from the full Lagrangian and applied the Dirac constraint formalism in order to obtain the Hamiltonian on the constraint surface. After appropriate redefinitions of the conjugate fields, it is possible to check that both  Hamiltonians are identical. For the sake of simplicity, here we only presented the more compact (but equivalent) derivation starting with the reduced Lagrangian.

\section{Discussion}
\label{sec:discussion}

In the first part of this final section, we emphasize that our results are compatible with a natural EFT perspective. In the second part, we discuss the phenomenology of the model and offer a brief outlook on future research.

\subsection{Effective field theory picture} \label{sec:EFT}

The EFT paradigm is based on the assumption that low energy physics can be decoupled from high energy physics. More precisely, if we are interested in describing a system at low energies, say below the scale $\Lambda$, we do not have to dynamically resolve degrees of freedom with masses greater than $\Lambda$. Instead, we write down an effective theory in which these heavy particles have been integrated out. This is achieved by including in the action all operators that respect the symmetries of the fundamental theory. In general, those operators come equipped with coefficients $c_i(\Lambda, \kappa)$ which depend on both the cut-off scale as well as the parameters (e.g.\ masses, coupling constants) of the fundamental theory, collectively denoted by $\kappa$. While the $\Lambda$-dependence is a relict of artificially introducing a cut-off and should be taken care of by an appropriate renormalization scheme, the $\kappa$-dependence is of physical relevance as it reflects the presence of the heavy degrees of freedom in nature. 

When we apply the EFT-reasoning to the BIG model, the action \eqref{eq:ActionBIG} should be regarded as the low energy version of a more fundamental theory of heavy particles that are localized on a four dimensional defect in a $ D $-dimensional bulk. By integrating out those heavy particles, the 4D curvature invariants denoted by $\mathcal{S}_{\rm BIG}$ are induced on the brane.  In this context, the presence of the induced gravity terms is a consequence of the low energy description and cannot be avoided in {\it any} brane setup. Since we know that there are consistent, i.e.\ ghost-free, microscopic theories describing localized particles~\cite{Dvali:2001ae}, there should also exist ghost-free versions of~\eqref{eq:ActionBIG} within a broad regime of induced parameter values. The last qualification is important since we require the parameters of the fundamental theory not to be fine-tuned\footnote{For the sake of simplicity we made the unnatural choice $\lambda^{(D)}=0$ for the bulk cosmological constant. However, we will see that it already suffices to choose a natural value for the brane tension $ \lambda $ in order to arrive at a ghost-free fluctuation theory.}. To be precise, we expect the process of integrating out heavy particles to generate induced operators with coefficients
\begin{align}\label{eq:EFT_ansatz}
	\lambda= c_1^4 M_*^4\;, && M_4=c_2 M_*\;, && r_0^{-1}=c_3 M_*\;,
\end{align}
where $c_1$,  $c_2$ and $c_3$ are dimensionless constants and $M_*$ is an arbitrary mass scale. We included $r_0$ because the brane width is ultimately also set by microscopic physics. Calculating the exact values for $c_i$ would of course require precise knowledge about the fundamental theory. 

For example, in the microscopic model discussed in \cite{Dvali:2001ae} the authors consider a localized scalar particle of mass $M$ that couples to the bulk gravity sector. On a perturbative level, it can be integrated out by calculating a particle loop on the brane. It is now crucial that this loop gives rise to both, an induced Einstein Hilbert term as well as a brane tension. Moreover, both terms are set by the same scale, implying $c_1 \sim c_2$. In particular, setting the tension to zero ($c_1=0$) while keeping $c_2$ nonzero would correspond to a fine-tuning, cf.\ Sec.~7 in \cite{Dvali:2001ae}.

Note that while, strictly speaking, the following discussion only applies to codimension two, we expect the qualitative results to be true in any higher codimension $n>1$.  When we plug \eqref{eq:EFT_ansatz} into \eqref{eq:def_f}, we obtain
\begin{equation}
	\label{eq:f_EFT}
	f = 1 - \frac{3 c_1^4 }{2 (c_2 c_3)^2} \frac{1}{1- \delta/(2 \pi)} \,,
\end{equation}
where the last factor may take any value in the range $[1,\infty)$.

The main result of Sec.~\ref{sec:lin_stab} showed that the theory is healthy, i.e.\ ghost-free, for parameters that obey $f<0$. Replacing the last factor in~\eqref{eq:f_EFT} by its greatest lower bound $ 1 $, we see that a sufficient condition for stability is given by $(c_2 c_3)^2/c_1^4<3/2$ (shown as a dotted line in Fig.~\ref{fig:cont}).
The important point is that this can always be achieved by choosing parameters of order one, i.e.\ without any fine-tuning. However, once we start to solely decrease the parameter $c_1$ (in~\cite{Berkhahn:2012wg} and Section~\ref{sec:hamiltonian} it was set to zero exactly), the bound gets quickly violated, cf.\ Fig.~\ref{fig:cont}. In other words, \textit{a consistent theory of gravity induced on a brane requires the inclusion of a sufficiently large brane tension}, which is in full accordance with an EFT perspective. 

\subsection{Phenomenology} \label{sec:pheno}
Independent of the question whether the stable regime can be realized without a fine-tuning of parameters, we may ask whether it is phenomenologically viable. To this end, let us derive the crossover scale of the model in 6D. It can be inferred directly from the brane-to-brane propagator of the healthy (observable) mode\footnote{Instead, we could also calculate the Newtonian potential for a point source which would contain a contribution from $s$, too. However, this would not change the conclusion.} $J|_0$ that is being convoluted with a constant point source $T_{00} \propto \delta^{3}(\vec{x})$. The correct expression for the propagator can be obtained from \eqref{eq:Z1} via the replacement $f \rightarrow -2$ [to see this compare \eqref{eq:healthy_mode} to \eqref{eq:israel_ghost}]. The time-independence of the source then implies $\omega=0$ and a crossover momentum $\vec{p}_c$ can be derived by comparing both terms in \eqref{eq:Z1}. Once they are of the same order, the induced graviton dynamics is as important as the bulk dynamics, thereby signaling a transition between a 4D and higher dimensional scaling behavior of the gravitational potential. 

To that end, we consider again Fig.~\ref{fig:rhs} which depicts the second term of \eqref{eq:Z1} as a function of $z= r_0 |\vec{p}|$. Now the crossover momentum can be derived by looking for intersections with the linear function $2 \alpha z$. The crossover momentum has to be much smaller than the inverse regularization scale, $r_0 |\vec{p}_c| \ll1$; otherwise, we would be sensitive to the unknown UV sector of the theory and---even more important---we would enter the higher dimensional regime already at microscopic distances below $r_0$ which is incompatible with observations. By a short inspection of Fig.~\ref{fig:rhs}, we see that this condition implies $\alpha \gg 1$ irrespective of the value of $\beta$. This means that only the right half of Fig.~\ref{fig:cont} is phenomenologically interesting. While in the sub-critical regime this is incompatible with having a stable theory, for near-critical values of the tension, i.e.\ $ 1-\delta/(2\pi)  \ll 1$, there is still a small stable stripe (also observed in \cite{Kaloper:2007ap}) between the stability line described by $f$ and the criticality line $\delta=2\pi$. A lower bound on the momentum crossover is found to be (neglecting factors of order one)  
\begin{align}
\label{eq:crossover}
	r_0|\vec{p}_c| \gtrsim \frac{1}{\alpha}	\qquad \text{for }\alpha \gg 1\,,
\end{align}
which leads in position space to an upper bound for the crossover distance $r_c \lesssim \alpha\, r_0$. This result matches the near-critical crossover scale derived in \cite{Kaloper:2007ap}.

In order to further assess the phenomenological status of the near-critical regime, we have to refer to cosmology. To be more specific, instead of considering a static (and 4D maximally symmetric background), we have to look for solutions with FLRW symmetries on the brane which allow for a non-vanishing Hubble parameter $H$ describing the expansion of the infinite brane dimensions. This was done on a fully nonlinear level in \cite{Niedermann:2014bqa}. As a result, a non-static generalization of the function $f$ (again separating a stable form an unstable regime) valid for FLRW symmetries was derived. The important point is that for $H\neq 0$ the stable near-critical parameter window becomes smaller. To be precise, it was found that for a stable, sub-critical solution to exist the following bound has to be fulfilled\footnote{The formula corresponds to Eq.~(50) in \cite{Niedermann:2014bqa}. Note that there a different ($\beta$-independent) definition for the crossover scale was used: $\check r_c^2=6 \alpha r_0^2$. In order to allow for the most conservative assessment, we express the same inequality in terms of the upper bound on the crossover derived here.}:
\begin{align}
	H r_c\lesssim 1\,.
\end{align}
This shows that the model does not allow for a stable 4D regime in cosmology (which would demand $H r_c \gg 1$). This agrees with the observation in~\cite{Niedermann:2014bqa} that the stable solutions are always governed by 6D dynamics. We therefore conclude that the sub-critical and near-critical model has to be dismissed in accordance with the analysis in~\cite{Niedermann:2014bqa}.

Let us stress that the present work and the cosmology analysis in \cite{Niedermann:2014bqa} complement each other very nicely. In both cases we find a stability line in parameter space, characterized by a function $f$, which separates a stable from an unstable regime. While in the cosmology analysis the qualification of a certain parameter regime as stable or unstable could only be inferred by solving the full differential system numerically, the present work permits an analytic assessment which shows perfect agreement for the case $H=0$.  On the other hand, only the cosmology analysis was able to rule out the near-critical parameter regime which still seemed to be viable for $H=0$.

Finally, let us comment on the range of applicability of the linear analysis presented here. Around static and spherically symmetric sources the codimension-one model is known to possess a Vainshtein-like radius below which the linear approximation breaks down. As it is parametrically large compared to the Schwarzschild radius, it turns out ot be crucial in restoring a 4D regime on solar system scales \cite{Deffayet:2001uk, Gruzinov:2001hp, Porrati:2002cp}. It is thus natural to ask whether a similar effect could exist in two codimensions as well. Answering this question would require to derive higher order corrections to the propagator in~\eqref{eq:Z1}, which is beyond the scope of the present work. Instead, we always assume to be in a weak coupling regime (which can be understood as a requirement on the source terms). However, since the (sub-critical) model is already ruled out by to the nonlinear analysis in~\cite{Niedermann:2014bqa}, which is obviously insensitive to that problem, this issue seems not relevant anymore.

\subsection{Outlook}
So far, it is not clear whether the picture might be more favorable in higher dimensions ($ D > 6 $). In Sec.~\ref{sec:hamiltonian} we showed that the instability is also present for arbitrary $ D > 6 $ on the parameter subspace $ \lambda = 0 $. But at the moment we do not know how the stability bound depends on the brane tension in more than two codimensions. This question is left for future research.

The remaining hope for the model in 6D resides on the super-critical regime. In a recent work (without BIG terms) it was shown that in this regime there is no stable Minkowski vacuum on the brane. Instead, the brane starts to expand in axial direction at a constant rate \cite{Niedermann:2014yka}. With respect to the cosmological constant problem, these solutions seem not to help since effectively the tension gets shifted by a constant amount ($\sim M_6^4$) which is by far to small to help with the problem.  Despite their incapability of addressing the naturalness question of the cosmological constant, they could nevertheless provide a viable infrared modification of gravity. We deem the super-critical regime therefore an exciting and promising playground for future research.

Another interesting questions concerns braneworld models with compact dimensions. This could be studied within the theory~\eqref{eq:ActionBIG} by simply imposing different boundary conditions. Consequently, the continuum of branch-cut states would be replaced by a discrete tower of massive Kaluza-Klein modes, see~\cite{Dvali:2001gm}. Since the ghost pole is independent from the cut, we suspect that the stability of the model would also be threatened in that context. However, an explicit calculation remains necessary.

Finally, let us recall that so far the bulk cosmological constant $ \lambda^{(D)} $ was always set to zero. It would be interesting to see how the healthy region in parameter space gets modified if also a natural value for $ \lambda^{(D)} $ of order $M_D^2$ was included.

\begin{acknowledgments}
We thank Felix Berkhahn, Gia Dvali, Fawad Hassan, Stefan Hofmann, Nemanja Kaloper and Justin Khoury for helpful discussions. FN would like to thank Claudia de Rham and Andrew Tolley for insisting on a calculation of the TT-amplitude, which partly motivated this work.
The work of FN was supported by TRR 33 ``The Dark Universe''.
The work of RS was supported by the DFG cluster of excellence ``Origin and Structure of the Universe''.
\end{acknowledgments}

\appendix

\section{Correcting the errors of the old ``no-ghost'' analysis}
\label{app:errata}

In the present work we have extended the ghost analysis of codimension-two BIG to a nontrivial brane-tension background. As discussed in the main text, the major result is that the ghost disappears if the tension is large enough, thus reconciling healthy codimension-two BIG with a natural EFT expectation.

For vanishing brane tension, however, the ghost always exists (if the induced gravity scale is nonzero). This result is in contradiction to those of Ref.~\cite{Berkhahn:2012wg}, where higher codimensional BIG was analyzed at the linear level around a 6D Minkowski background, and claimed to be stable. In this appendix, we resolve this tension by explicitly identifying the errors made therein.
\subsection{Lorentz covariant analysis} 
\label{app:cov_analysis}
One claim in \cite{Berkhahn:2012wg} was that the scalar $ \mathcal{S} $ (see below)---which corresponds to the ghost mode identified in earlier works (as well as in the present one)---does not threaten the stability of the model because it would be a constrained quantity. We will now show that this is wrong: $ \mathcal{S} $ is in fact not constrained, but dynamical.

When we linearize the theory on a 6D Minkowski background, i.e.\ for a vanishing deficit angle, the full dynamical system \eqref{eq:vac_einstein} and \eqref{eq:israel} can be written in a concise way by introducing $\delta$--functions. In Cartesian coordinates, we find
\begin{multline}
	\label{app:einstein1}
	M_D^{D-2}\Diamond_{MN}^{(D)\;AB}h_{AB}=\\
	\mixInd{\delta}{\mu}{M} \mixInd{\delta}{\nu}{N}\,\delta_n(y)\left(\tilde U_{\mu\nu}-M_4^2  \Diamond_{\mu\nu}^{(4)\;\rho\sigma}h_{\rho\sigma}\right)\,.
\end{multline}
where $\Diamond_{MN}^{(D)\;AB}$ is the first order Einstein operator in $D$ dimensions
\begin{multline}
	\Diamond_{MN}^{(D)\;AB} =  \mixInd{\delta}{A}{M} \mixInd{\delta}{B}{N}\, \Box_D + \eta^{AB}\partial_M \partial_N - 2\, \mixInd{\delta}{B}{\!\!(N}\, \partial^A  \partial_{M)}\\
						+ \eta_{MN}\left(\partial^A \partial^B - \eta^{AB}\, \Box_D \right)\,,
\end{multline}
{and the 4D localized source $\tilde U_{\mu\nu}$ is related to the first order energy momentum tensor via $\tilde U_{\mu\nu}:= -2\,\T^{(4)\alpha}_{\hphantom{(4)\mu}\nu} $.}
We decompose the graviton field in a 4D Lorentz covariant way,
\begin{subequations}\label{app:decomp1}
	\begin{align}
		h_{\mu\nu}
		&=\mathcal{D}_{\mu\nu} + \partial_{(\mu} \mathcal{V}_{\nu)} +
		\partial_{\mu}\partial_{\nu} \; \mathcal{B} + \eta_{\mu\nu} \; \mathcal{S}
		\,,\\
		h_{\mu b}
		&=\mathcal{E}_{\mu b} + \partial_\mu \mathcal{F}_{b} +
		\partial_{b}\mathcal{G}_\mu + \partial_{\mu}\partial_{b} \mathcal{H}
		\,.
	\end{align}
\end{subequations}
The purely extradimensional components $ h_{ab} $ are decomposed as in \eqref{eq:split_ab}. Here, $\mathcal{D}_{\mu\nu}$ is a transverse and traceless 4D tensor, $\mathcal{V}_{\mu}$, $\mathcal{E}_{\mu b}$, $\mathcal{G}_{\mu}$  are transverse 4D vectors and $\mathcal{B}$, $\mathcal{S}$, $\mathcal{F}_b$, $\mathcal{H}$ as well as all functions appearing in \eqref{eq:split_ab} are 4D scalars. Furthermore, $\mathcal{F}_b$ and $\mathcal{E}_{\mu b}$ transform as vectors under the SO($n$) group. 

Even though this decomposition makes the Lorentz covariance of the dynamical equations manifest, it has the general disadvantage that the components are not determined uniquely.  More precisely, this ambiguity can be parametrized in terms of a set of homogeneous functions $\chi^{(i)}$, where here and henceforth ``homogeneous'' refers to solutions of the 4D homogeneous wave equation, i.e.\ $\Box_4 \chi^{(i)}=0$. It can be easily checked that the decomposition \eqref{app:decomp1} is then invariant under the transformations
\begin{subequations}\label{app:split_amb1}
	\begin{align}
		\delta \mathcal{S}			&= \chi^{(1)}\,, \label{app:amb1}\\ 
		\delta \mathcal{B}			&= \chi^{(2)}  -\frac{4}{\Box_4}\chi^{(1)}\,,\\
		\delta \mathcal{V}_{\mu}		&=  \chi^{(3)}_{\mu} + \frac{3}{\Box_4} \partial_{\mu}\chi^{(1)} \,,\label{app:amb2}\\
		\delta \mathcal{D}_{\mu\nu}	& = \left[ 4\,\partial_{\mu}\partial_{\nu} \frac{1}{\Box_4} - 3\, \partial_{(\mu}\frac{1}{\Box_4}\partial_{\nu)} - \eta_{\mu\nu} \right]\chi^{(1)} \nonumber\\
								&\quad - \partial_\mu\partial_\nu \chi^{(2)} - \partial_{(\mu}\chi^{(3)}_{\nu)}\,, \label{app:amb_D}
	\end{align}
\end{subequations}
as well as
\begin{subequations}\label{app:split_amb2}
	\begin{align}
		\delta \mathcal{H}			&= \chi^{(4)}\,,				& \delta \mathcal{F}_{a}		&= \chi^{(5)}_a\,, \label{app:amb3}\\
		\delta \mathcal{G}_{\mu}		&= -\partial_{\mu}\chi^{(4)}\,,	& \delta \mathcal{E}_{\mu b}	&= -\partial_{\mu}\chi^{(5)}_b\,,
	\end{align}
\end{subequations}
where $\chi^{(3)}_{\mu}$ and $\chi^{(5)}_{a}$ are subject to the two conditions
\begin{align} \label{app:amb_restrict}
	\partial^\mu \chi^{(3)}_{\mu} = -3 \partial^{\mu}\frac{1}{\Box_4}\partial_{\mu} \chi^{(1)} \,, && \partial^a \chi^{(5)}_{a}=0 \,.
\end{align}
Here and henceforth, $(1/\Box_4) \psi$ is a shorthand notation for the convolution of $\psi$ with the retarded Green's function of the 4D d'Alembert operator $\Box_4$.

Besides the split ambiguity, there is the usual gauge freedom \eqref{eq:gauge}.
%, which decomposes into
%%
%\begin{subequations}
%\begin{gather}
%	\delta \mathcal{B} = \frac{2}{\Box_4} \partial^\mu \xi_\mu \quad \delta b = \frac{2}{\Delta_2} \partial^a \xi_a \\
%	\delta H = \frac{1}{\Box_4} \partial^\mu \xi_\mu + \frac{1}{\Delta_2} \partial^a \xi_a \\
%	\delta \mathcal{V}_\mu = \xi_\mu - \partial_\mu\frac{1}{\Box_4}\partial^\nu \xi_\nu = -\delta \mathcal{G}_\mu \\
%	\delta v_a = \xi_a - \partial_a\frac{1}{\Delta_2}\partial^b \xi_b = -\delta \mathcal{F}_a
%\end{gather}
%\end{subequations}
Instead of fixing a particular gauge, we will again use a complete set of gauge invariant variables, viz.\ $ \mathcal{S}, s, \mathcal{D}_{\mu\nu}, \mathcal{E}_{\mu b} $ and $ d_{ab} $, as well as
\begin{subequations}\label{app:gauge_inv}
	\begin{align}
		\mathcal{O}		&:=\mathcal{B}+b-2 \mathcal{H}\,,\label{app:C}\\
		\mathcal{W}_{\mu}	&:=\mathcal{G}_{\mu}-\mathcal{V}_{\mu}\,,\label{app:inv2}\\
		\mathcal{Y}_{a} 	&:=\mathcal{F}_{a} - v_{a}\,.\label{app:inv3}
	\end{align}
\end{subequations}
Although these are invariant under gauge transformations, they are not invariant under the homogeneous transformations \eqref{app:split_amb1} and \eqref{app:split_amb2}, but transform as
\begin{subequations}
	\begin{align}
		\delta \mathcal{O} & =  -\frac{4}{\Box_4}\chi^{(1)} + \chi^{(2)} - 2\chi^{(4)} \,, \label{app:amb_C}\\
		\delta \mathcal{W}_{\mu} & = - \frac{3}{\Box_4} \partial_{\mu}\chi^{(1)} - \chi^{(3)}_{\mu} - \partial_{\mu}\chi^{(4)} \,, \label{app:amb_W}\\
		\delta \mathcal{Y}_{a} & = \chi^{(5)}_a \,.
	\end{align}
\end{subequations}
Now it is crucial to realize that one certain combination of the functions $ \chi^{(2)}, \chi^{(3)}_\mu $ and $ \chi^{(4)} $ does not affect the gauge invariant quantities. Explicitly, one can easily check that they only enter via the combinations\footnote{Note that $ \tilde\chi^{(3)}_\mu $ is subject to the same relation~\eqref{app:amb_restrict} as $ \chi^{(3)}_\mu $, and thus also only has three independent components.}
\begin{align}
	\tilde\chi^{(2)} := \chi^{(2)} - 2 \chi^{(4)} \,, && \tilde\chi^{(3)}_\mu := \chi^{(3)}_\mu + \partial_\mu \chi^{(4)} \,,
\end{align}
(or any linear combination thereof) in the relevant Eqs.~\eqref{app:amb_D}, \eqref{app:amb_C} and \eqref{app:amb_W}.

In summary, the gauge invariant variables $ \{ \mathcal{S}, s, \mathcal{O}, \mathcal{W}_\mu, \mathcal{Y}_a, \mathcal{D}_{\mu\nu}, \mathcal{E}_{\mu b}, d_{ab}  \} $ are only unique up to the $ 4 + n $ independent homogeneous functions $\{\chi^{(1)}, \tilde\chi^{(2)}, \tilde\chi^{(3)}_{\mu}, \chi^{(5)}_{a}\}$. This will be crucial for correctly inferring the number of dynamical degrees of freedom (DOF), because each of the $ \chi $'s can be used to eliminate one would-be dynamical component.

To derive the equations of motion for our variables, we start by investigating the $\mu b$-components of \eqref{app:einstein1}. Taking their double divergence $\partial_{\mu}\partial_{b}$ leads to 
\begin{equation}
	\Box_4 \Delta_n \left(3\, \mathcal{S} + (n-1) s \right)=0\;.
\end{equation}
Demanding fall-off conditions at spatial infinity allows to simply drop the extra space Laplace operator $\Delta_n$. The general solution then becomes $3\, \mathcal{S} + (n-1) s =\kappa^{(s)}$ with $\kappa^{(s)}$  an arbitrary homogeneous function. Instead of choosing initial conditions to fix  $\kappa^{(s)}$, we make use of the split ambiguity \eqref{app:amb1} parametrized by $\chi^{(1)}$ to remove it from the equation. $s$ is therefore constrained by the relation 
\begin{equation}\label{app:eq_s}
	s =-\frac{3}{n-1}\mathcal{S}\;,
\end{equation}
and therefore no independent DOF.

By acting with a single divergence $\partial_{\mu}$ or $\partial_{b}$ on the $\mu b$-components of  \eqref{app:einstein1} and using~\eqref{app:eq_s}, we obtain % equations for the gauge invariant combinations \eqref{app:inv3} and \eqref{app:inv2}, respectively. After dropping the overall spatial Laplace operators and using~\eqref{app:eq_s}, these equations are again simple wave equations,
\begin{align}
	\Box_4 \Delta_n \mathcal{W}_\mu = 0 \,, && \Box_4 \Delta_n \mathcal{Y}_a = 0 \,,
\end{align}
respectively. As before, $ \Delta_n $ can be dropped, and the freedom to choose initial conditions gets ``eaten'' by $\tilde\chi^{(3)}_{\mu}$ and $\chi^{(5)}_{a}$, yielding
\begin{equation}
	\mathcal{W}_{\mu} = \mathcal{Y}_{a} = 0 \,.
\end{equation}
These relations can be used to simplify the $\mu b$-sector of \eqref{app:einstein1}, leading to the wave equation
\begin{equation}
	\Box_D \mathcal{E}_{\mu b}=0 \,.
\end{equation}
Thus, the fields $\mathcal{E}_{\mu b}$ constitute $3(n-1)$ DOF that are not sourced by brane-localized matter fields. 

There is only one freedom left in choosing the decomposition, namely the function $ \tilde\chi^{(2)} $. We will use it in the same way as before to derive a constraint equation for $\mathcal{O}$. After taking the trace of the $ab$-components of \eqref{app:einstein1} and using \eqref{app:eq_s}, we find
\begin{align}
	\Box_4 \mathcal{O}=-\frac{n-2}{n-1}\mathcal{S}\;.
\end{align}
A priori, this is a dynamical equation for $\mathcal{O}$ sourced by $\mathcal{S}$. However, according to \eqref{app:amb_C} the decomposition is invariant under the shift $\delta \mathcal{O}= \tilde\chi^{(2)}$. Again, this implies that $\mathcal{O}$ is not a true DOF as we can impose arbitrary initial conditions without affecting the solution for $h_{AB}$. For instance, we can choose $\tilde\chi^{(2)}$ such that $\mathcal{O}$ becomes 
\begin{align}
	\mathcal{O}=-\frac{n-2}{n-1} \frac{1}{\Box_4} \mathcal{S}\,.
\end{align}
Once we plug this solution back into the $ab$-components of \eqref{app:einstein1}, we find a wave equation for the  transverse and traceless SO($n$)-tensor modes,
\begin{equation}
	\Box_D d_{ab}=0 \,, 
\end{equation}
which therefore constitute further $ [(n+1)(n-1)/2] $ DOF that are not coupled to on-brane matter.

Finally, let us consider the $\mu\nu$-components of \eqref{app:einstein1}. Taking its trace and making use of all solutions we derived before yields an equation for the scalar $\mathcal{S}$,
\begin{align}\label{app:ghost}
	M_D^{D-2}\,\Box_D \mathcal{S}=\frac{1}{3}\frac{n-1}{n+2}\left(\mixInd{\tilde U}{\mu}{\mu} + 6 M_4^2\,\Box_4 \mathcal{S}\right) \delta_{n}(y).
\end{align}

Since we already made use of all shift ambiguities, $\mathcal{S}$ is a real dynamical mode. In fact, for $n=2$ it coincides with the tachyonic ghost mode from the analysis in Sec.~\ref{sec:lin_stab} and hence fulfills the same dynamical equation. In order to demonstrate the equivalence, we use \eqref{app:eq_s} to rewrite \eqref{app:ghost} in terms of $s$ and use the regularization introduced in Sec.~\ref{sec:regularization} by simply replacing 
\begin{align}
	\delta_2(y) \rightarrow \frac{\delta(r-r_0)}{2\pi r_0}\,.
\end{align}
Integrating the resulting equation over the interval $[r_0-\epsilon, r_0+\epsilon]$ yields the matching equation \eqref{eq:israel_ghost} with $\delta=0$ and $\mixInd{U}{\mu}{\mu}=\mixInd{\tilde U}{\mu}{\mu}/(2\pi r_0)$; the vacuum equation \eqref{eq:bulk_scalar_dyn} is obtained by evaluating the equation for $r \neq r_0$. 

Having thus established the ghost character of $\mathcal{S}$ (or $s$ equivalently) for $n=2$, we turn to the remaining sourced DOF in the general case. In order to derive the dynamical equation for the 4D tensor $\mathcal{D}_{\mu\nu}$, we
need to solve for $\mathcal{S}$ more explicitly. To be precise, $\mathcal{S}$ fulfills the equation 
\begin{multline}\label{app:ghost2}
	M_D^{D-2}\,\left(\frac{\Delta_n}{\Box_4}+1 \right) \mathcal{S} = \\
	\frac{1}{3}\frac{n-1}{n+2}\Bigg(\frac{1}{\Box_4}\mixInd{\tilde U}{(4)\mu}{\!\mu}
	 +6M_4^2\, \mathcal{S}\Bigg)\delta_n(y) +\kappa^{(\mathcal{S})}
	\,,
\end{multline}
where we introduced the homogeneous function $ \kappa^{(\mathcal{S})} $, which keeps track of the freedom to choose initial conditions for $\mathcal{S}$. (Recall that there is no split ambiguity left which could be used to set $ \kappa^{(\mathcal{S})} $ to zero. This subtlety was missed in~\cite{Berkhahn:2012wg}, cf.\ the discussion below.) This allows to derive an equation for the tensor~$\mathcal{D}_{\mu\nu}$,
\begin{multline} \label{app:D}
	M_D^{D-2}\,\Box_D \mathcal{D}_{\mu\nu}=\left(\tilde U^{(\mathcal{D})}_{\mu\nu} -M_4^2\, \Box_4 \mathcal{D}_{\mu\nu}\right)\delta_n(y)\\
	+\partial_{\mu}\partial_{\nu}\kappa^{(\mathcal{S})}
	\,,
\end{multline}
with the transverse and traceless tensor $\tilde U^{(\mathcal{D})}_{\mu\nu}$ defined as
\begin{align}
	\tilde U^{(\mathcal{D})}_{\mu\nu} := \tilde U_{\mu\nu} + \frac{\partial_{\mu}\partial_{\nu}}{3}\frac{1}{\Box_4} \mixInd{\tilde U}{\rho}{\rho} - \frac{1}{3}\eta_{\mu\nu} \mixInd{\tilde U}{\rho}{\rho}\,.
\end{align}
This equation is not completely decoupled, since it depends on the initial conditions for $\mathcal{S}$ through $\kappa^{(\mathcal{S})}$. According to the analysis in Sec.~\ref{sec:lin_stab}, the sign of the induced term does not imply a ghost. As $\mathcal{D}_{\mu\nu}$ is also subject to the five constraint equations $ \mixInd{\mathcal{D}}{\mu}{\mu} = 0 $ and $\partial^{\mu}\mathcal{D}_{\mu\nu}=0$, it describes five healthy DOF. Therefore, the theory contains 6 sourced $\{\mathcal{S}, \mathcal{D}_{\mu\nu}\}$ and $[(3+n/2)(n-1)-1]$ non-sourced $\{\mathcal{E}_{\mu b}, d_{ab}\}$ DOF, which makes a total of $D (D-3)/2$, corresponding to the number of DOF in $D$-dimensional GR, in accordance with the results of the Hamiltonian analysis in Sec.~\ref{sec:hamiltonian}. 

To summarize, the manifestly covariant analysis exactly recovers the results of the analysis in Sec.~\ref{sec:lin_stab} and generalizes it to arbitrary higher codimensions. For $n=2$, the five healthy DOF can be identified according to  $\mathcal{D}_{\mu\nu}\rightarrow \{D_{ij}, C_{i}, J\}$, whereas in both cases the ghost is described by $\mathcal{S}$ (or $s$ equivalently).

In the remainder of this appendix, we want to point out the error in the analysis of Ref.~\cite{Berkhahn:2012wg}. There, the identical dynamical equation~\eqref{app:ghost} for $\mathcal{S}$ was found, indicating that $\mathcal{S}$ is a ghost. It was then argued that there would be a further, more restrictive \emph{constraint} equation for $ \mathcal{S} $ which could be derived from the (00)-component of the modified Einstein equations, rendering  $ \mathcal{S} $ non-dynamical. For its derivation it was necessary to use of the dynamical equation for $\mathcal{D}_{\mu\nu}$.  However, that equation in Ref.~\cite{Berkhahn:2012wg} differs from \eqref{app:D} by the last term $\partial_{\mu}\partial_{\nu}\kappa^{(\mathcal{S})}$. The reason is that the equation was derived by applying the transverse-traceless projector $\mathcal{O}^{(4, \rm tt)\rho\sigma}_{\;\;\;\mu\nu}$ (for a definition see the appendix of~\cite{Berkhahn:2012wg}) to the equations of motion~\eqref{app:einstein1} and tacitly assuming that it commutes with the linearized Einstein operator $\Diamond^{(D)\;\rho\sigma}_{\mu\nu}$. However, in general the commutator of those operators is non-zero but yields a homogeneous function due to the occurrence of the Green's function $1/\Box_4$ in the projector. This fact caused the failure of the analysis in~\cite{Berkhahn:2012wg}.

More explicitly, for an arbitrary function $\psi$ the following relation can be derived easily:
\begin{equation}
	\left[\frac{1}{\Box_4}, \partial_\mu\right]\psi = \kappa_{\mu}[\psi] \,,
\end{equation}
where $ \kappa_\mu $ are four homogeneous functions that are uniquely determined by $ \psi  $. This expression could be generalized to the commutator of the projector $\mathcal{O}^{(4, \rm tt)\rho\sigma}_{\;\;\;\mu\nu}$ with $\Diamond^{(D)\;\rho\sigma}_{\mu\nu}$. Instead of applying the projection operator explicitly, we decided to derive~\eqref{app:D} by consecutively solving the scalar and vector sector and keeping track of all homogeneous functions as described above. By doing so, we avoid all potential pitfalls related to the non-commutativity of the projector. Finally, it is straightforward to check that the (00)-Einstein equation is identically fulfilled and does not provide any new informations; in particular, it is no constraint on $\mathcal{S}$. More explicitly, we can eliminate all fields except for $\mathcal{S}$ from it and arrive at an expression which is equivalent to equation \eqref{app:ghost2}. However, due to the occurrence of the homogeneous function $\kappa^{(\mathcal{S})}$ (which was missed in the old analysis), this should not be interpreted as a constraint on $ \mathcal{S} $, it is simply a different (but equivalent) formulation of the dynamical equation \eqref{app:ghost}.

\subsection{Hamiltonian analysis} 
\label{app:hamiltonian}

The analysis of Sec.~\ref{sec:hamiltonian} showed that the Hamiltonian on a Minkowski background for $n>1$ is not bounded from below, which is another manifestation of the ghost mode $s$. This is another clear contradiction to the result in~\cite{Berkhahn:2012wg}, where it was claimed that the Hamiltonian is positive definite for $n=2$. To resolve it, we have to reconsider the gauge choices in combination with the symmetry assumptions that were made in~\cite{Berkhahn:2012wg}.

Beside the choice $\Pi_i^i 	=0$, the following two gauges were imposed, cf.\ Eqs.~(47)--(49)\footnote{All equation numbers here refer to the Journal version of Ref.~\cite{Berkhahn:2012wg}.} in~\cite{Berkhahn:2012wg},
\begin{align}
\partial_i h_{ij}=0\,, &&  h_{56}=0\,.
\end{align}
In order to check whether these gauge choices are consistent and leave any residual gauge freedom, we have to infer their transformation behavior according to \eqref{eq:gauge},
\begin{align}
\delta_{\xi} \left(\partial_i h_{ij}\right) 	&= \frac{1}{2}\left( \Delta_3 \xi_j + \partial_j \partial_i \xi_i \right)\,,\\
\delta_{\xi} h_{56} 					&= \partial_{(5}\xi_{6)}\;.
\end{align} 
It follows that the first gauge completely fixes $\xi_i$ without any residual gauge freedom. The second gauge leaves a residual gauge freedom in the $\xi_a$-sector which is subject to the condition $\partial_{(5}\xi_{6)} = 0$. Furthermore, the gauge functions $\xi_A$ have to respect the SO(2) symmetry that was also assumed in~\cite{Berkhahn:2012wg}. Explicitly, this requires $\xi_\phi=0$ and $\partial_{\phi} \xi_r = 0$, which in Cartesian coordinates implies $\partial_{[5}\xi_{6]}=0$. This relation together with the residual gauge condition completely fixes the two functions $\xi_a$ and does not allow for any further gauges. In particular, it is no longer possible to gauge the combination $2\partial_r \partial_i h_{5i}- \cos(\phi) (\Delta_{3}+\Delta_{2})h_{55}$ as it was done in Eq.~(59) in~\cite{Berkhahn:2012wg}. However, this gauge was crucial in that case to demonstrate the positive definiteness of the Hamiltonian. Correspondingly, the corrected constraint analysis of Sec.~\ref{sec:hamiltonian} comes to a different result.

\section{Junction conditions} \label{ap:junct_cond}

Here, we present the derivation of the junction conditions~\eqref{eq:israel_tensor}--\eqref{eq:israel_phi} in more detail.
A straight-forward calculation yields the following non-vanishing components of the extrinsic curvature tensor $ \mixInd{K}{\alpha}{\beta} $ at linear order around the deficit angle background~\eqref{eq:met_bkg}:
\begin{subequations}
\begin{align}
	\prescript{1}{}{K}^\mu_{\hphantom{\mu}\nu} & = \frac{1}{2} \left.\left( \partial_r \mixInd{h}{\mu}{\nu} - \partial^{\mu} h_{\nu r} - \partial_{\nu} \mixInd{h}{\mu}{r} \right) \right|_0 \,,\\
	\prescript{1}{}{K}^\phi_{\hphantom{\phi}\phi} & = \frac{1}{2} \left.\left( \partial_r \mixInd{h}{\phi}{\phi} - \frac{g'}{g} \mixInd{h}{r}{r} \right ) \right|_0 \,.
\end{align}
\end{subequations}
The junction conditions are all obtained by plugging this (and the linearized 5D Einstein tensor) into~\eqref{eq:israel}, projecting onto the desired components, and simplifying the BIG terms by means of the bulk equations in the limit $ r \to r_0 $.
For the tensor $ (ij)^{(D)} $ and vector $ (0i)^{(N)} $ components, this immediately gives the junction conditions~\eqref{eq:israel_tensor} and~\eqref{eq:israel_vector}.
Similarly, the two scalar junction conditions~\eqref{eq:israel_scalar} are obtained by taking the required linear combinations of the 4D trace, scalar $ (ij)^{(S)} $ and $ (\phi\phi) $ equations.

Next, we have to consider the $ (\phi\phi) $ component. After using the jump of the $ r $-derivative of the bulk equation~\eqref{eq:bulk_scalar_3}, it takes the simple form
\begin{equation}
%	M_6^4 \left (  [J'] + \frac{\delta}{2\pi r_0} d \right ) + M_5^3 \Box_4 J|_0 = U_{00} - \Delta_3 & U^{(B)} \,,\\
%%	M_6^4 \left (  [J'] + \Delta_3 [B'] + \frac{\delta}{2\pi r_0} d \right ) & \nonumber\\
%%	+ M_5^3 \Delta_3 \left (2 S + d\right )|_0 & = U_{00} \,,\\
%%	M_6^4 [B'] + M_5^3 \left.\left ( S + d - Q \right)\right|_0 & = U^{(B)} \,,\\
%	-M_6^4 \left ( [N'] + \Delta_3[B'] + 2[S'] + [s'] +  \frac{\delta}{2\pi r_0} d \right ) & \nonumber\\
%	+ M_5^3 \left.\left ( \Delta_3 Q + 2 \ddot S - \Delta_3 S  - \Box_4 d \right)\right|_0 & = U^{(S)} \,,\\
%	-M_6^4 \left ( [N'] + \Delta_3[B'] + 3[S'] \right ) + M_5^3 \Box_4 s|_0 = U^\phi_{\hphantom{\phi}\phi} \label{eq:israel_phi_ap} \,.
	M_6^4 \Box_4 [b'] + M_5^3 \Box_4 s|_0 = U^\phi_{\hphantom{\phi}\phi} \label{eq:israel_phi_ap} \,.
\end{equation}
%
%Again, the bulk equations in the limit $ r \to r_0 $ were used to simplify the BIG term. It can be further simplified by noting that the jump of the $ r $-derivative of the bulk equation~\eqref{eq:bulk_scalar_3}, together with continuity of the metric (implying $ [H'] = [l'] = 0 $) yields:
%%
%\begin{equation} \label{eq:jump1}
%	[N'] + \Delta_3[B'] + 3[S'] + \Box_4 [b'] = 0 \,.
%\end{equation}
%%
Furthermore, the second equation in~\eqref{eq:h_b_s} shows that continuity of $ \mixInd{h}{\phi}{\phi} $ implies
\begin{equation} \label{eq:jump}
	[b'] = \left[ \frac{g}{g'} \right] \left (\varphi - s|_0 \right) = \frac{r_0 \delta}{2\pi - \delta} \left (\varphi - s|_0 \right) \,.
\end{equation}
Using~\eqref{eq:jump} in~\eqref{eq:israel_phi_ap} yields Eq.~\eqref{eq:israel_phi}.

%Note that all other junction conditions are redundant due to the Gauss-Codazzi (and bulk vacuum) equations.
For completeness, let us note that the jump in the $ r $-derivative of the constrained scalar $ O $ is determined by the $ (ij)^{(B)} $ component of Israel's junction conditions. After using again the bulk equations, as well as~\eqref{eq:jump} and the fact that $ [H'] = 0 $ (due to continuity of the metric), it simplifies to
\begin{multline}\label{eq:jump_c}
	M_6^4 \left\{ [O'] + \frac{r_0 \delta}{2\pi - \delta} \left(s|_0 - \varphi \right) \right \} \\
	+ M_5^3 \left.\left( \frac{J - s}{3} - Q + \varphi \right )\right|_0 = U^{(B)} \,,
\end{multline}
where an overall $ \Delta_3 $ was dropped.
The jumps in the $ r $-derivative of all the remaining constrained gauge-invariant quantities ($ W_i, P $ and $ Q $) can readily be obtained from the $ r $-derivatives of the corresponding bulk equations.

As a consistency check, we explicitly verified that all the junction conditions, together with the bulk equations, imply the energy conservation equations~\eqref{eq:en_cons}, as is guaranteed by the Gauss-Codazzi equations.

\section{Non-stabilized circumference} \label{ap:non_stabil}

In this appendix, we investigate the case when the brane circumference is not kept constant. This simply means that the radion $ \varphi $ is not set to zero, and $ \mixInd{U}{\phi}{\phi} $ is arbitrary. For simplicity, we will only consider the case
\begin{equation}
	\mixInd{U}{\phi}{\phi} = 0 \,.
\end{equation} 
The two junction conditions~\eqref{eq:israel_s} and~\eqref{eq:israel_phi} can still be used to derive a closed equation\footnote{Note that when comparing this equation to the one derived in Ref.~\cite{Kaloper:2007ap}, viz.~Eq.~(5.41) therein (the corresponding scalar mode is called $ X $ in~\cite{Kaloper:2007ap}, and is related to ours via $ s = -3X $), one has to take into account that the energy momentum tensor $ \mixInd{\tau}{\mu}{\nu} $ in~\cite{Kaloper:2007ap} differs from our $ \mixInd{U}{\mu}{\nu} $: $ \mixInd{\tau}{\mu}{\nu} $ is defined in the Einstein frame and thus satisfies standard energy conservation ($ \partial_\mu \mixInd{\tau}{\mu}{\nu} = 0 $), whereas $ \mixInd{U}{\mu}{\nu} $ is defined in the Jordan frame, for which, on the deficit angle background, additional terms $ \propto \lambda $ have to be included, cf.\ our Eq.~\eqref{eq:en_cons}. We thank Nemanja Kaloper for clarifying this point.} for $ s $,
\begin{multline} \label{eq:israel_s_non_stabil}
	M_6^4 [s'] + M_5^3 \left[ \left( 1 - \frac{3\alpha}{2\beta} \right) \Box_4 + \frac{\beta - 2\alpha}{2\alpha \left( 1 + \beta \right)r_0^2} \right] s|_0 \\
	= - \frac{1}{4} \mixInd{U}{\mu}{\mu} \,. %+ \left[ \frac{3}{4} \left( 1 - \frac{2\alpha}{\beta} \right) - \frac{1}{(1 + \beta)r_0^2 \Box_4} \right] \mixInd{U}{\phi}{\phi} \,. 
\end{multline}
After performing a 4D Fourier transform and using the general bulk solution~\eqref{eq:s_bulk} for $ \hat s $, we arrive at
\begin{equation}
	\frac{4M_6^4}{r_0} \tilde Z(p) \hat s|_0 = - \hat U^\mu_{\hphantom{\mu}\mu} \,,
\end{equation}
where now the inverse $ s $-propagator is given by (in terms of $ z := r_0 \sqrt{p^2} $)
\begin{equation}
	\tilde Z(p) := \alpha f_1 z^2 + f_2 - z \left ( \frac{I_1(z)}{I_0(z)} + \frac{K_1((1 + \beta) z)}{K_0((1 + \beta) z)}  \right ) \,,
\end{equation}
with
\begin{align}
	f_1 := 2 \left( \frac{3\alpha}{2\beta} - 1 \right) \,, && f_2 := \frac{\beta - 2\alpha}{1 + \beta} \,.
\end{align}
This is very similar to the inverse $ s $-propagator in the stabilized case, Eq.~\eqref{eq:Z1}. The only difference is that the coefficient $ f $ is slightly modified into $f_1 $ and---more importantly, as we will see below---there is an additional constant (i.e.~$ p $-independent) term $ f_2 $.

As before, we can restrict ourselves to sources for which only the scalar mode $ s $ is excited. This can be achieved by setting all tensor- and vector source terms to zero, and requiring
\begin{equation}
	3\hat U^{(S)} = \left( 1 - \frac{f_2}{4\tilde Z} \right) \mixInd{\hat U}{\mu}{\mu} \,.
\end{equation}
In this case, the scalar mode $ J $ can again be set to zero, and the full source vertex~\eqref{eq:def_A} takes the same form as before~\eqref{eq:s_vertex}, with the replacement $ Z \to \tilde Z $. Therefore, the stability analysis is completely analogous.

\begin{figure}[htb]
	\centering
		\includegraphics[width=0.45\textwidth]{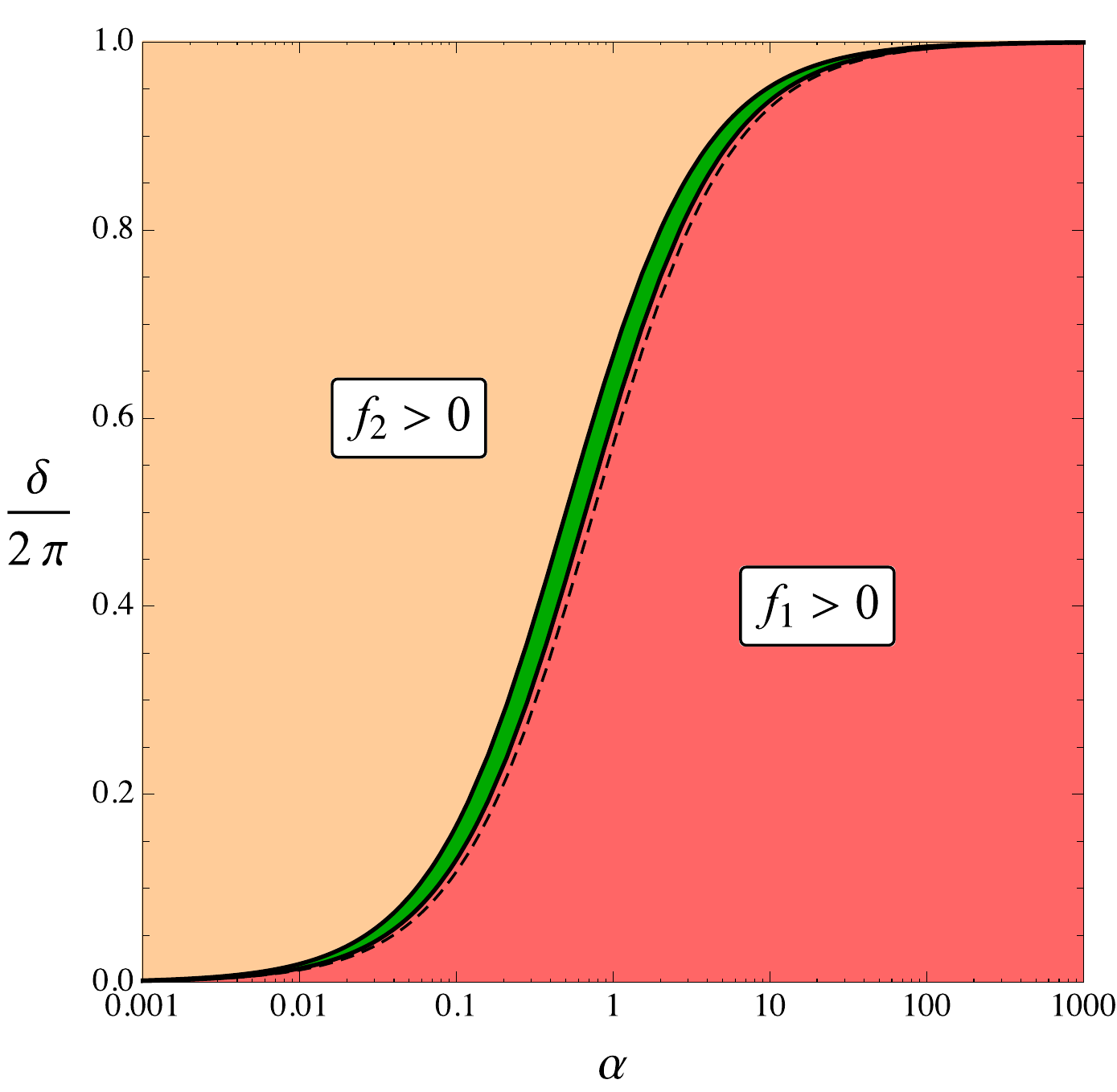}
		\caption{Without stabilization (and $ U^{\phi}_{\;\;\phi} = 0 $), there are three regions in parameter space: in the lower right region, where $ f_1 > 0 $, there is a tachyonic ghost; the delimiting line is almost the same as in the non-stabilized case, shown as a dashed line. In the dark green region in the middle $ f_1, f_2 < 0 $, and the model is linearly stable. In the upper left region $ f_2 > 0 $ and there is a tachyon which is not a ghost. It shows that the static deficit angle background is not stable if $ U^{\phi}_{\;\;\phi} $ is not used to fix the brane circumference.}
		\label{fig:cont_non_stabil}
\end{figure}

It can now easily be checked that the propagator $ \tilde Z^{-1} $ has a tachyonic pole ($ \Leftrightarrow \tilde Z$ is zero for some $ z_* > 0 $) if and only if $ f_1 > 0 $ or $ f_2 > 0 $. The corresponding regions in parameter space are depicted in Fig.~\ref{fig:cont_non_stabil}. They are disjoint, and separated by a narrow (but finite) stripe in which the tachyon is absent and the model is thus linearly stable.

The case $ f_1 > 0 $ is very similar to the condition $ f > 0 $ in the stabilized case. Indeed, the delimiting line $ f = 0 $ in parameter space, shown as a dashed line in Fig.~\ref{fig:cont_non_stabil}, only gets shifted by a small amount. 

The case $ f_2 > 0 $, however, implies that there is also a tachyon in the upper left region in parameter space. At first, it might be surprising that this region is also unstable, because it also includes the case $ \alpha = 0 $, which is just pure 6D GR without any induced terms and should be a healthy theory. The resolution to this puzzle is rather simple: Evaluating the residue of $ \tilde Z^{-1} $ at the tachyon pole as we did in Sec.~\ref{sec:tach_ghost}, we find that, while it is again negative for $ f_1 > 0 $, it is \emph{positive} for $ f_2 > 0 $. In other words, the tachyon is only a ghost in the lower right region. In the upper left region, the tachyon is not a pathology, but merely a reflection of the fact that, \textit{without fixing the brane circumference, the static deficit angle background is not stable.} Instead, the brane wants to expand (or collapse) in radial direction.

To summarize, the ghost-criterion is basically independent of whether the circumference is stabilized or not. In particular, the naturalness discussion from Sec.~\ref{sec:EFT} still applies to the case of free radial expansion. On the other hand, the static deficit angle solution is not stable under angular size fluctuations. To overcome this problem, we have to make additional assumptions about the underlying microscopic model. In fact, from a fundamental perspective, the existence of some sort of stabilization mechanism has to be expected as there are known stable vortex configurations in two codimensions, cf.~\cite{Nielsen:1973cs}. Fixing the proper circumference turned out to be a convenient way of realizing such a mechanism in an effective low energy description. In the following appendix we study an explicit microscopic example.

\section{Example of a stabilization mechanism} \label{ap:radion_stabil}

In this appendix, we will present an explicit realization of a stabilization mechanism, which is capable of keeping the circumference of the ring-regularized brane (approximately) constant. In the main text, it was assumed that some mechanism of this kind could exist, and so it is instructive to see a concrete example. Following~\cite{Kaloper:2007ap}, we put a massless scalar field $ \Sigma $ on the brane, i.e.~we add
\begin{equation}
	S_\Sigma = - \int \!\rd^5 x \, \sqrt{-g^{(5)}} \frac{1}{2} \partial^\alpha \Sigma \partial_\alpha \Sigma
\end{equation}
to the action~\eqref{reg_sub}. We then choose a background solution for which the scalar field winds around the ring,
\begin{equation}
	\Sigma = q \phi \,,
\end{equation}
where $ q $ is a constant. Together with~\eqref{reg_sub}, this leads to the following background energy momentum tensor (EMT),
\begin{subequations} \label{eq:EMT_axion}
	\begin{align}
	\mixInd{T}{(5)\mu}{\nu} &= -  \left(M_5^3 \lambda^{(5)} + \frac{q^2}{4} g^{\phi\phi} \right) \delta^\mu_\nu \,, \label{eq:Tmunu_axion}\\
	\mixInd{T}{(5)\phi}{\phi} &= - M_5^3 \lambda^{(5)} + \frac{q^2}{4} g^{\phi\phi} \,,
	\end{align}
\end{subequations}
where $ g^{\phi\phi} = 1/r_0^2 $ at the background level. As discussed in Sec.~\ref{sec:def_angle_sol}, the static pure tension solution exists for $ \mixInd{T}{(5)\phi}{\phi} = 0 $, which can now be achieved by choosing $ q $ such that
\begin{equation}
	\frac{q^2}{4r_0^2} = M_5^3 \lambda^{(5)} \,.
\end{equation}
From~\eqref{eq:Tmunu_axion} we see that this doubles the contribution of $ \lambda^{(5)} $ to the 4D EMT. After absorbing this factor by a trivial renormalization\footnote{In the main text, we work with the renormalized quantity $ \lambda^{(5)}_\mathrm{(ren)} $.}, $  \lambda^{(5)}_\mathrm{(ren)} \equiv 2 \lambda^{(5)}_\mathrm{(bare)} $, and identifying the 4D brane tension via
\begin{equation}
	\lambda \equiv 2\pi r_0 M_5^3 \lambda^{(5)}_\mathrm{(ren)} \,,
\end{equation}
we arrive at the background EMT~\eqref{eq:EMT_4D_tension}.

So winding $ \Sigma $ around the ring in this way indeed leads to the static deficit angle solution presented in Sec.~\ref{sec:def_angle_sol}. But a successful stabilization should also suppress the fluctuations of the brane circumference, as measured by the radion field $ \varphi \equiv h_{\phi\phi} $. Let us now show that this can also be achieved in this particular example.

First, note that the fluctuations of $ \Sigma $ can be consistently set to zero, since this field is not sourced.
Next, perturbing the background metric in~\eqref{eq:EMT_axion} leads to the following first order contribution to the EMT,
\begin{equation}
	\mixInd{U}{\phi}{\phi} = \frac{M_6^4}{r_0} \frac{\delta}{2\pi} \varphi \,,
\end{equation}
where we eliminated $ \lambda^{(5)} $ in favor of the deficit angel $ \delta $. [There are similar contributions to $ \mixInd{U}{\mu}{\nu} $, which cancel the corresponding $ \varphi $-terms on the left hand side of~\eqref{eq:israel_scalar}.] Plugging this into~\eqref{eq:israel_phi}, it becomes a mass term for the radion,
\begin{align}
	\left( \Box_4 - m_\varphi^2 \right) \varphi + \left ( 1 - \frac{2\alpha}{\beta}\right ) s|_0 = 0 \,, && m_\varphi^2 := \frac{1 - \delta/2\pi}{r_0^2} \,.
\end{align}
Since $ m_\varphi \propto 1 / r_0 $, the radion gets heavier as the regularization size $ r_0 $ decreases, and so it costs more energy to excite this DOF. In the low energy regime we are aiming at, i.e.~at energies well below $ 1/r_0 $, the kinetic term of $ \varphi $ is negligible,%
\footnote{This is not true in the near-critical regime, where the factor $ 1 - \delta/2\pi $ suppresses the radion mass. However, our main results about the ghost mode are not affected by this, because they even hold without any stabilization, as shown in Appendix~\ref{ap:non_stabil}. Furthermore, we are here only discussing one particular stabilization mechanism. There might also be other examples, where the brane width is also stabilized in the near-critical regime, in which case the analysis in the main text applies there as well.}%
and we obtain
\begin{equation}
	\varphi \approx \left (\frac{2\alpha}{\beta} - 1 \right ) \frac{\Box_4}{m_\varphi^2} s|_0 \,.
\end{equation}
The remaining explicit $ \varphi $-term in~\eqref{eq:israel_s} is thus negligible compared to $ s|_0 $ and can be dropped, while $ \mixInd{U}{\phi}{\phi} $ can be replaced by
\begin{equation}
	\mixInd{U}{\phi}{\phi} \approx \left (M_5^3 - \beta M_6^4 r_0 \right ) \Box_4 s|_0 \,.
\end{equation}
This is exactly the same as~\eqref{eq:Uphiphi}, and we indeed recover the scalar equations~\eqref{eq:israel_scalar_stabil} from the main text. While there, they were derived by assuming some underlying stabilization mechanism, and then inferring the required $ \mixInd{U}{\phi}{\phi} $ from the field equations, we have now seen an explicit example that gives the radion a large mass, and consistently yields the same $ \mixInd{U}{\phi}{\phi} $ and field equations after the radion has been integrated out.

\newpage
\bibliography{Ghostbusters_v3}

\end{document}